\pdfoutput=1
\documentclass[twocolumn, showpacs,amsmath,amssymb,aps,prl,floatfix,longbibliography, 10pt]{revtex4-1}

\usepackage[latin1]{inputenc}
\usepackage[american]{babel}
\usepackage[T1]{fontenc}
\usepackage{graphicx}
\usepackage{mathrsfs}
\usepackage{hyperref}
\usepackage{epstopdf}
\usepackage{bm, epsfig}

\hypersetup{colorlinks=true, urlcolor=blue, citecolor=blue, filecolor=blue, linkcolor=blue}
\newcommand*{\diff}{\mathop{}\!\mathrm{d}}

\newcommand*{\Imm}{\mathop{}\!\mathbf{Im}}
\newcommand{\uimm}{\mathrm{i}}
\newcommand{\eu}{\mathrm{e}}
\newcommand{\diag}{\mathrm{diag}}
\newcommand{\daga}{^{\dagger}}

\begin{document}

\title{Deterministic strong-field quantum control}

\author{Stefano~M.~Cavaletto}
\email[Email: ]{smcavaletto@gmail.com}
\affiliation{Max-Planck-Institut f\"ur Kernphysik, Saupfercheckweg 1, 69117 Heidelberg, Germany}
\author{Zolt\'an~Harman}
\affiliation{Max-Planck-Institut f\"ur Kernphysik, Saupfercheckweg 1, 69117 Heidelberg, Germany}
\author{Thomas~Pfeifer}
\affiliation{Max-Planck-Institut f\"ur Kernphysik, Saupfercheckweg 1, 69117 Heidelberg, Germany}
\author{Christoph~H.~Keitel}
\affiliation{Max-Planck-Institut f\"ur Kernphysik, Saupfercheckweg 1, 69117 Heidelberg, Germany}
\date{\today}
\begin{abstract}
Strong-field quantum-state control is investigated, taking advantage of the full---amplitude and phase---characterization of the interaction between matter and intense ultrashort pulses via transient-absorption spectroscopy. A sequence of intense delayed pulses is used, whose parameters are tailored to steer the system into a desired quantum state. We show how to experimentally enable this optimization by retrieving all quantum features of the light-matter interaction from observable spectra. This provides a full characterization of the action of strong fields on the atomic system, including the dependence upon possibly unknown pulse properties and atomic structures. Precision and robustness of the scheme are tested, in the presence of surrounding atomic levels influencing the system's dynamics.
\end{abstract}


\maketitle

{The advent of laser light and femtosecond pulse-shaping technology have revolutionized our access to the quantum properties of matter \cite{1367-2630-12-7-075008, tannor2007introduction, PhysRevA.37.4950}, with coherent-control methods exploiting interference in order to steer a system into a given state with light} \cite{doi:10.1146/annurev.pc.43.100192.001353, PhysRevLett.68.1500, MeshulachNature, WeinachtNature, BrixnerNature}. 
Measurement-driven techniques {such as adaptive feedback control} are extensively used, especially when little understanding of the light-matter interaction is available owing to inaccurately known atomic or molecular structures, nonideal experimental conditions, or because of the use of strong, insufficiently characterized laser fields. Femtosecond pulses are thus utilized to simultaneously control and interrogate the atomic system, with their shape being iteratively optimized based on the received experimental response \cite{PhysRevLett.68.1500}. However, the associated atomic dynamics remain concealed in the optimal pulse, often preventing insight into the underlying physical mechanism. {Only recently techniques were investigated to access the complex reaction pathways followed by an optimally controlled system \cite{Daniel536, PhysRevLett.110.223601} and in the strong-field regime, where perturbative approaches fail and the atomic level structure is dressed by the time-dependent field, a limited number of effective pulse-shaping strategies has been identified \cite{PhysRevLett.94.083002, PhysRevLett.100.233603, PhysRevLett.102.023004, PhysRevA.81.063410}.}

{Major advances in x-ray free-electron lasers (FELs) are now enabling quantum control also at short wavelengths \cite{RevModPhys.88.015006}. 
Coherent transform-limited x-ray pulses are produced via seeding methods at FELs \cite{nphoton.2012.180, AllariaNaturePhot}, opening the field of x-ray quantum optics \cite{doi:10.1080/09500340.2012.752113}. Despite 
recent advances \cite{PhysRevLett.115.114801, PrinceNaturePhot}, however, experimental challenges still need be faced. Complex spectral-shaping methods are not yet available at short wavelengths, in particular for hard x~rays, and control schemes, e.g., to manipulate several excited states lying within the x-ray pulse bandwidth, should preferably rely on optimal pulse sequences. Methods to measure pulse temporal profiles are significantly hindered at x-ray frequencies by the absence of suitable nonlinear crystals. Therefore, measurement-driven strategies directly accessing the atomic response to intense, insufficiently characterized pulses should be preferred to methods based on theoretical assumptions of the pulse shape. At the same time, the reduced flexibility at recently established x-ray FELs renders adaptive feedback still very challenging for current experiments.}


\begin{figure}[t]
\includegraphics[width= \columnwidth]{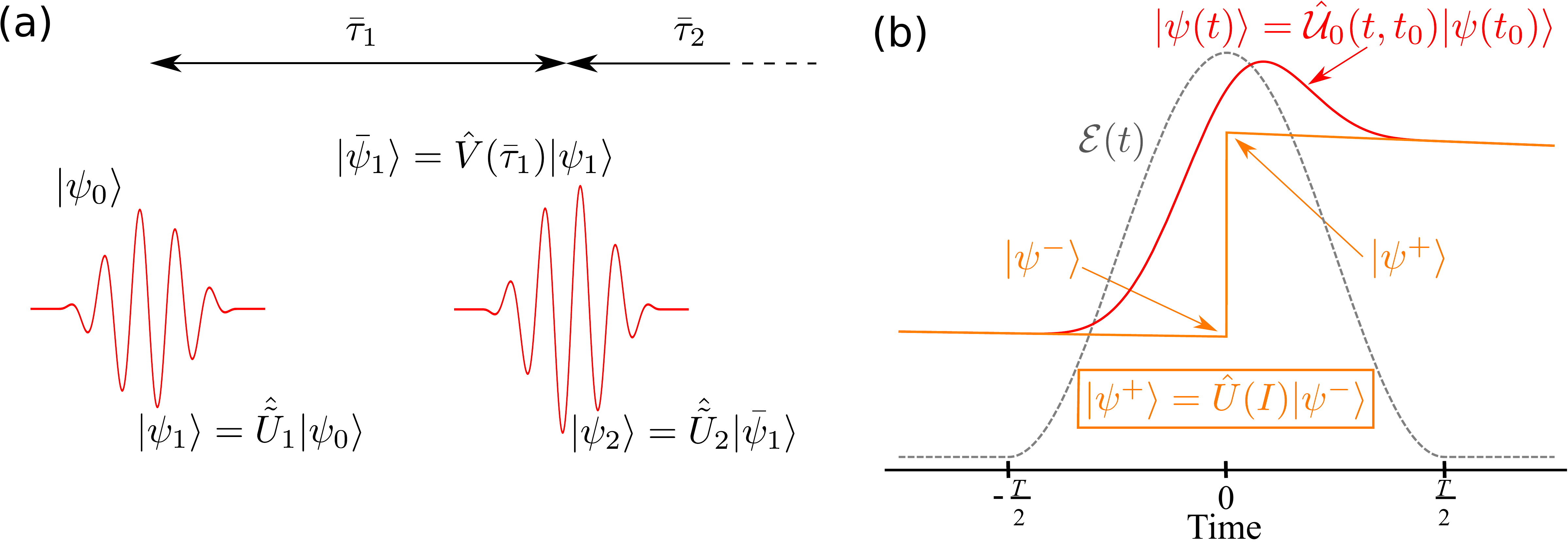}
\caption{(Color online). (a) Quantum-control scheme based on an optimal sequence of experimentally characterized pulses. (b) Interaction of a quantum system (red, continuous line) with an ultrashort pulse (gray, dashed line), effectively modeled as an instantaneous effect (orange, continuous line).}
\label{fig:intro}
\end{figure}

{In order to determine an effective route to x-ray quantum control despite present limitations, in this Letter we put forward a scheme to experimentally characterize---in amplitude and phase---the atomic interaction with intense ultrashort pulses, and use this information to deterministically guide the system into a desired state with an optimal pulse sequence [Fig.~\ref{fig:intro}(a)]. Thereby, one keeps the advantages of a measurement-driven strategy. In stark contrast to adaptive feedback, however, where optimal pulse shapes are iteratively determined via a trial-and-error procedure, our scheme allows access and visualization of the building blocks constituting the optimal strong-field control strategy, providing an advantageous means to unravel the dynamical pathways followed by the system. Furthermore, for experimental conditions in which feedback may not be advisable due to, e.g., restricted beam time, our scheme represents a cost-effective strategy to prepare a given system in different states: adaptive feedback requires a new sequence of iterations for every desired state, whereas deterministic strong-field control could be performed numerically relying on a set of elementary steps characterized experimentally. 
Motivated by recent results \cite{PhysRevLett.115.033003}, the scheme is applied here to control optical transitions in Rb atoms, but it could be straightforwardly implemented at x-ray energies with, e.g., highly charged ions, among the best candidates for future x-ray quantum-optics applications \cite{doi:10.1080/09500340.2012.752113}.}



{\it Interaction operators.}
The key quantity we will use to characterize the atomic response to intense ultrashort pulses is the interaction operator $\hat{U}(I)$, whose action is represented in Fig.~\ref{fig:intro}(b). 
We assume a pulse of the form $\boldsymbol{\mathcal{E}}(t) =  \mathcal{E}_0\,f\left(t\right)\,\cos\left(\omega_{\mathrm{L}} t\right)\hat{\boldsymbol{e}}_z$, linearly polarized along the $\hat{\boldsymbol{e}}_z$ unit vector, with laser frequency $\omega_{\mathrm{L}} = 1.59\,\mathrm{eV}$ and peak field strength $\mathcal{E}_0 = \sqrt{8\pi\alpha I}$, where $I$ is the pulse peak intensity and $\alpha$ the fine-structure constant \cite{diels2006ultrashort}. The envelope function $f(t)$ is nonvanishing in the interval $[-T/2,\, T/2]$, with pulse duration $T$. Atomic units are used unless stated otherwise. In the absence of external fields, the known free evolution of the system under the action of the atomic structure Hamiltonian $\hat{H}_0$ is given by $\hat{V}(t) = \eu^{-\uimm\hat{H}_0 t}$. In the interval $[-T/2,\, T/2]$, however, the dynamics of the system $|\psi(t)\rangle = \hat{\mathcal{U}}_0(t,t_0)|\psi(t_0)\rangle$, depicted in Fig.~\ref{fig:intro}(b), require the solution of the Schr\"odinger equation in the presence of the external pulse $\boldsymbol{\mathcal{E}}(t)$. In order to operatively describe this strong-field interaction, we 
introduce the {\it effective} initial and final states $|\psi^{\mp}\rangle = \eu^{\uimm\hat{H}_0(\mp T/2)}|\psi(\mp T/2)\rangle$, represented in Fig.~\ref{fig:intro}(b). The unique, intensity-dependent operator $\hat{U}(I) = \eu^{-\uimm \hat{H}_0 (-T/2)}\hat{\mathcal{U}}_0\left(T/2,-T/2\right)\eu^{\uimm \hat{H}_0 (T/2)}$, connecting $|\psi^+\rangle$ with $|\psi^{-}\rangle$,
\begin{equation}
|\psi^+\rangle=\hat{U}(I)|\psi^-\rangle,
\label{eq:U}
\end{equation}
is used to effectively describe the action of an ultrashort pulse in terms of a $\delta$-like interaction \cite{PhysRevA.83.033405, 0953-4075-47-12-124008}. 

{Endowed with an efficient way to quantify the action of strong ultrashort pulses, we can summarize our deterministic quantum-control scheme as follows. To prepare a system in a desired state $|\psi_{\mathrm{d}}\rangle$, we use the sequence of $N_{\mathrm{p}}$ pulses shown in Fig.~\ref{fig:intro}(a), separated by delays $\bar{\tau}_m$ and leading the system to the state
\begin{equation}
|\psi_{N_{\mathrm{p}}}\rangle = \hat{\tilde{U}}_{N_{\mathrm{p}}} \ldots \hat{V}(\bar{\tau}_m)\hat{\tilde{U}}_m \ldots \hat{V}(\bar{\tau}_1)\hat{\tilde{U}}_1 |\psi_{\mathrm{0}}\rangle.
\label{eq:control1}
\end{equation}
Here, the action of the $m$th pulse is described by $\hat{\tilde{U}}_m = \hat{\varPhi}\daga(\phi_m)\,\hat{U}(I_m)\,\hat{\varPhi}(\phi_m)$, with intensity $I_m$ and where $\hat{\varPhi}(\phi_m)$ accounts for a carrier-envelope phase (CEP) $\phi_m$. In adaptive feedback, a series of experiments is performed for every desired state $|\psi_{\mathrm{d}}\rangle$, iteratively searching for the optimal combinations of pulse delays, CEPs, and intensities. Little knowledge is thereby achieved about the possible pathways the system could follow and the rules determining the optimal pulse sequence. In contrast, in deterministic strong-field control, experiments are first run to fully characterize the interaction operators $\hat{U}(I)$, providing a complete experimental mapping of the available control options and facilitating manipulation and interpretation of the chosen control strategy.} Although interaction operators could be calculated from theory, our deterministic scheme allows one to effectively tackle those cases where reliable predictions are not possible via methods based exclusively on theory, due to missing knowledge of the atomic structures, pulse shapes, or the strong-field interaction.



\begin{figure}[t]
\includegraphics[width= \columnwidth]{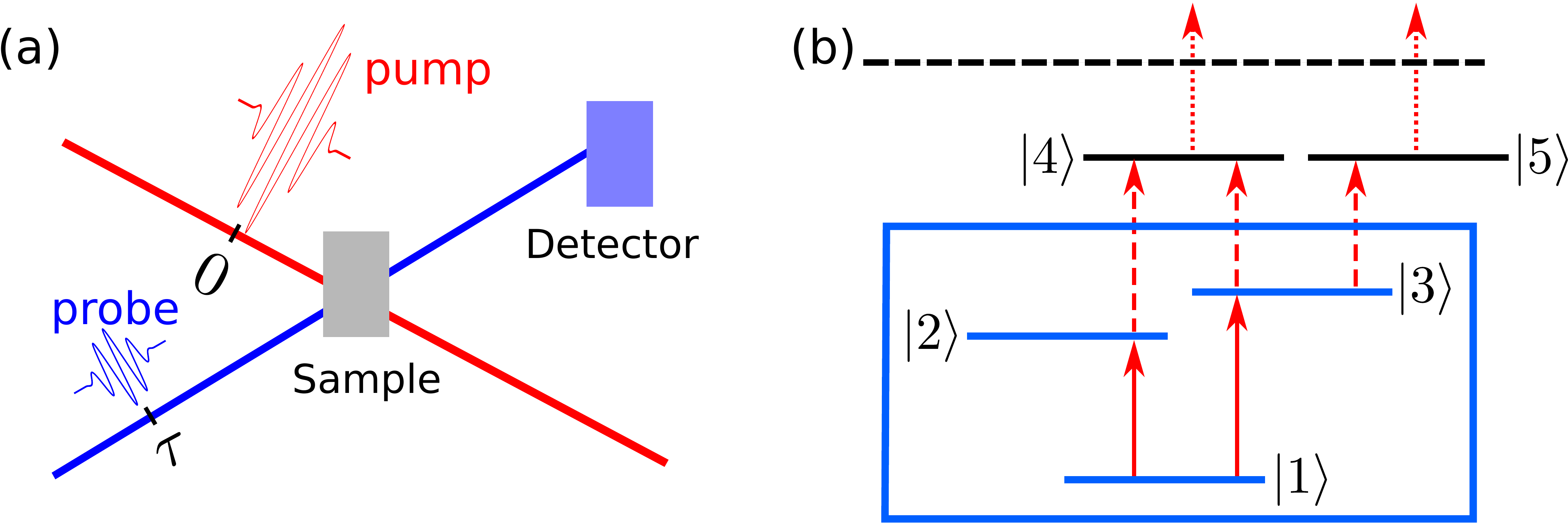}
\caption{(Color online). (a) Transient-absorption-spectroscopy setup used to experimentally reconstruct strong-field interaction operators. (b) Level scheme used to model Rb atoms, aiming at the control of the $V$-type three-level scheme in the box.}
\label{fig:model}
\end{figure}


{\it Reconstruction of the strong-field interaction operators.} {To exploit the advantages of a measurement-driven strategy without employing adaptive feedback,} we utilize transient-absorption spectroscopy (TAS) to reconstruct $\hat{U}(I)$ in amplitude and phase, and use these extracted matrices for quantum-state control. TAS has been receiving increasing interest for studies of ultrafast dynamics \cite{Mathies06051988, doi:10.1146/annurev.pc.43.100192.002433, PhysRevLett.98.143601, PhysRevLett.105.143002, GoulielmakisNature466, PhysRevLett.106.123601, Ott10052013, 0953-4075-49-6-062003, Meyer22122015}. In a pump-probe setup, depicted in Fig.~\ref{fig:model}(a), the absorption spectrum $\mathcal{S}(\omega,\tau)$ of a transmitted weak probe pulse is observed for varying time delays $\tau$, revealing the dynamics initiated by the intense pump pulse \cite{RevModPhys.40.441}. At the same time, recent experiments have employed a probe-pump scheme ($\tau<0$), with the probe pulse generating a coherent superposition of quantum states which is subsequently nonlinearly excited by the strong pulse \cite{PhysRevLett.105.143002, PhysRevA.86.063408, 1367-2630-16-11-113016, PhysRevLett.112.103001}. In this case, absorption spectral line shapes contain valuable information to quantify the strong-field dynamics induced by the pump pulse, albeit requiring schemes to extract information from complex time-dependent spectra. {Characterizing strong-field interactions to reconstruct $\hat{U}(I)$ with TAS can be straightforwardly implemented experimentally, since the same intense pulse is used with varying time delays. This minimizes the number of experiments where pulse parameters need be precisely modified, in contrast to adaptive feedback, where pulse intensities, phases and delays are simultaneously and controllably varied {\it at every iteration} to converge to the desired state.}

We apply our scheme to Rb atoms \cite{PhysRevA.65.043406, PhysRevLett.115.033003}. Specifically, we aim at controlling the $V$-type three-level system formed by the ground state $5s\,^2S_{1/2}\equiv|1\rangle$ and fine-structure-split excited states $5p\,^2P_{1/2}\equiv|2\rangle$ and $5p\,^2P_{3/2}\equiv|3\rangle$, with magnetic quantum numbers $M = \pm 1/2$ and transition energies $\omega_{21} = 1.56\,\mathrm{eV}$ and $\omega_{31} = 1.59\,\mathrm{eV}$. The electric-dipole-($E1$-)allowed transitions $|1\rangle\rightarrow|k\rangle$, $k\in\{2,\,3\}$ [box in Fig.~\ref{fig:model}(b)] feature $\Delta M = 0$ and dipole-moment matrix elements $\boldsymbol{D}_{1k} = D_{1k} \hat{\boldsymbol{e}}_z$ \cite{PhysRevA.30.2881}. {The decay of the system is accounted for in the atomic structure Hamiltonian $\hat{H}_0 = \sum_i{(\omega_{i} -\uimm\gamma_i/2 ) |i \rangle\langle i|}$, while the interaction with femtosecond pump and probe pulses, respectively centered on $t = 0$ and $t=\tau$, is included via the $E1$ interaction Hamiltonian in the rotating-wave approximation \cite{Scully:QuantumOptics}.} $\gamma_{k}$ are set to $1/(1500\,\mathrm{fs})$ to model experimental finite linewidths due to, e.g., Doppler and collision-induced broadening. {The much smaller spontaneous-decay rates, of $\sim \mathrm{ns}^{-1}$, can be neglected for the femtosecond time scales of interest. A Schr\"odinger formalism is used to describe the state of the system, whereas a more thorough treatment based on density matrices is presented in the Supplemental Information.} 

\begin{figure}[t]
\includegraphics[clip=true, width=\columnwidth]{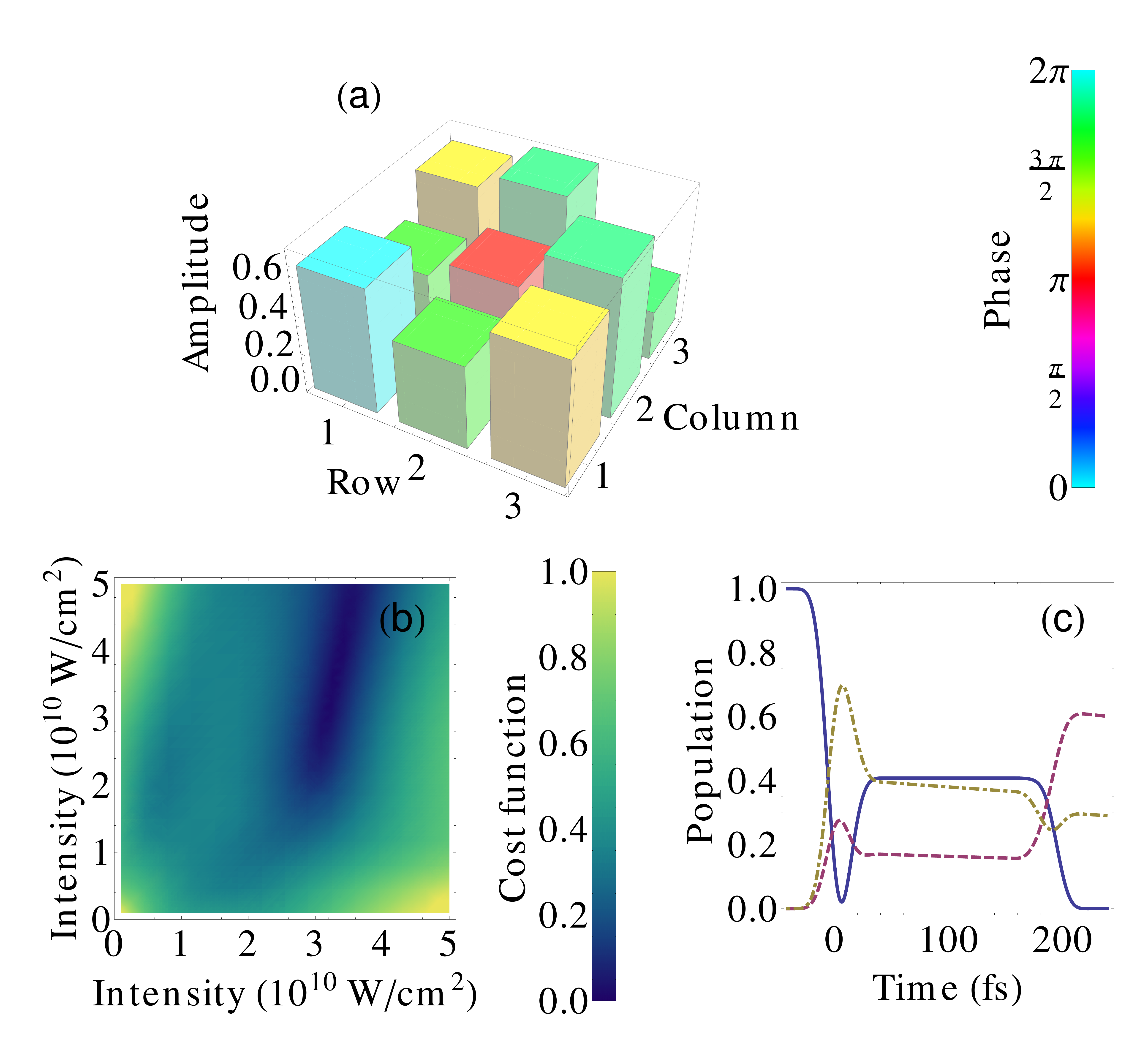}
\caption{(Color online). (a) Reconstructed SFI operator $\hat{U}^{\mathrm{R}}(I)$ for $I = 3.3\times 10^{10}\,\mathrm{W/cm^2}$, with bar heights (colors) exhibiting matrix-element amplitudes (phases). [(b)-(c)] Two-pulse scheme implemented for $I_1 = 3.3\times10^{10}\,\mathrm{W/cm^2}$, $I_2 = 3.6\times10^{10}\,\mathrm{W/cm^2}$, $\bar{\tau} = 198\,\mathrm{fs}$, and $\bar{\phi} = 1.88\,\mathrm{rad}$, with (b) a section of the control landscape as a function of $I_1$ and $I_2$, for fixed $\bar{\tau}$ and $\bar{\phi}$ and (c) the corresponding evolution of the populations of state $|1\rangle$ (blue, continuous), $|2\rangle$ (purple, dashed), and $|3\rangle$ (yellow, dotdashed), with a reached final state $|\psi_{\mathrm{r}}\rangle = \sum_{i=1}^3 c_{\mathrm{r},i}|i\rangle$ featuring $|c_{\mathrm{r},2}|^2/|c_{\mathrm{r},3}|^2 = 2.07$ and $c_{\mathrm{r},1} = 0.00$.}
\label{fig:twopulses}
\end{figure}

For low densities, {where propagation effects can be neglected and the pulses can be assumed to homogeneously control the sample,} the time-delay-dependent spectra result from the system's single-particle dipole response,
\begin{equation}
\mathcal{S}(\omega,\tau)\propto -\omega\Imm\left[\hat{\boldsymbol{e}}_z\cdot \int_{-\infty}^{\infty}\left\langle\hat{\boldsymbol{D}}^{-}(t,\tau)\right\rangle\,\eu^{-\uimm \omega (t- \tau)}\diff t\right],
\label{eq:spectrum}
\end{equation}
where $\langle\hat{\boldsymbol{D}}^{-}(t,\tau)\rangle = \left\langle\psi(t,\tau)\right|\hat{\boldsymbol{D}}^{-}\left|\psi(t,\tau)\right\rangle$ is the expectation value of the dipole-moment operator $\hat{\boldsymbol{D}}^{-} = \sum_{i>j} \boldsymbol{D}_{ij} |i \rangle\langle j|$, and $|\psi(t,\tau)\rangle$ describes the time evolution of the system as a function of time delay. We use Eq.~(\ref{eq:spectrum}) to numerically simulate experimental spectra $\mathcal{S}_{\mathrm{exp}}(\omega,\tau)$ for pump pulses of intensities varying between $0.1\times10^{10}\,\mathrm{W/cm^2}$ and $5\times10^{10}\,\mathrm{W/cm^2}$ and for the noncollinear geometry of Fig.~\ref{fig:model}(a) \cite{PhysRevLett.115.033003}. This is based on the full solution of the Schr\"odinger equation for a three-level system interacting with delayed pump and probe pulses. At the same time, Eq.~(\ref{eq:spectrum}) is also used to derive a fitting model for transient-absorption spectra and, thereby, enable the extraction of strong-field interaction (SFI) operators. For this purpose, we take advantage of the same instantaneous-interaction model introduced in Eq.~(\ref{eq:U}) and formally describe the system's dynamics in a pump-probe experiment in terms of the operators $\hat{U}_{\mathrm{pr}}$ and $\hat{U}_{\mathrm{pu}}(I)$. For weak probe pulses, first-order perturbation theory is used to model $\hat{U}_{\mathrm{pr}}$, whereas the matrix elements $U_{\mathrm{pu},ij}$ of the intensity-dependent pump-pulse operator $\hat{U}_{\mathrm{pu}}(I)$ are unknown fitting parameters. For $\tau<0$, the effective evolution of the time-delay-dependent state $\left|\psi_{\mathrm{fit}}(t,\tau)\right\rangle$ from the initial state $|\psi_0\rangle = |1\rangle$ can be modeled as
\begin{equation}
|\psi_{\mathrm{fit}}(t,\tau)\rangle = \left\{
\begin{aligned}
&|\psi_0\rangle, &\text{if $t<\tau$},\\
&\hat{V}(t-\tau)\hat{U}_{\mathrm{pr}}|\psi_0\rangle, &\text{if $\tau<t<0$},\\
&\hat{V}(t)\hat{U}_{\mathrm{pu}}(I)\,\hat{V}(-\tau)\,\hat{U}_{\mathrm{pr}}|\psi_0\rangle, &\text{if $t>0$},
\end{aligned}
\right.
\label{eq:evolutionprobepump}
\end{equation}
with analogous formulas for $\tau>0$. Inserting this in Eq.~(\ref{eq:spectrum}), an analytical fitting model $\mathcal{S}_{\mathrm{fit}}(\omega,\tau,U_{\mathrm{pu},ij})$ can be derived (see details in the Supplemental Information) and used to fit the experimental spectra $\mathcal{S}_{\mathrm{exp}}(\omega,\tau)$ and reconstruct the SFI operators $\hat{U}^{\mathrm{R}}(I)$ in amplitude and phase:
\begin{equation}
\left.
\begin{aligned}
\mathcal{S}_{\mathrm{exp}}(\omega,\tau)\\
\mathcal{S}_{\mathrm{fit}}(\omega,\tau,U_{\mathrm{pu},ij})
\end{aligned}
\right\} \underset{\text{fitting}}{\longrightarrow} 
\text{Reconstructed operator $\hat{U}^{\mathrm{R}}$}.
\label{eq:UR}
\end{equation}
The effectiveness of the method is exemplified in Fig.~\ref{fig:twopulses}(a), where we display the extracted SFI matrix $\hat{U}^{\mathrm{R}}(I)$ for a pump intensity of $I = 3.3\times 10^{10}\,\mathrm{W/cm^2}$. 
The same reconstruction scheme could be implemented in an experiment, enabling access to strong-field light-matter interactions without requiring knowledge of pump-pulse intensities or the system's dynamics.

{\it Quantum control guided by experimentally characterized pulses.} Once SFI operators are reconstructed as a function of pulse intensities, these are employed to implement our deterministic control method from Eq.~(\ref{eq:control1}). In the following, we focus on a two-pulse scheme, and use reconstructed SFI operators to optimize time separation $\bar{\tau}$, intensities $I_{m}$, and CEPs $\phi_m$, $m\in\{1,\,2\}$, to control the populations of the final state $|\psi_{2}\rangle$. This yields a predicted final state 
\begin{equation}
|\psi_{\mathrm{p}}\rangle = {\textstyle \sum_{i = 1}^3 }c_{\mathrm{p},i}|i\rangle =\hat{U}^{\mathrm{R}}(I_2)\,\hat{\varPhi}(\bar{\phi})\,\hat{W}(\bar{\tau})\,\hat{U}^{\mathrm{R}}(I_1)\,|1\rangle,
\label{eq:sequenza}
\end{equation}
where we neglect phase terms not influencing the final populations, and introduce the total phase $\bar{\phi} = \phi_2 - \phi_1 - \omega_{\mathrm{L}}\bar{\tau}$, the CEP operator $\hat{\varPhi}(\phi) = \diag(1,\,\eu^{\uimm\phi},\,\eu^{\uimm\phi})$, and the slowly oscillating operator $\hat{W}(\bar{\tau}) = \diag\left(1,\,\eu^{-[\gamma_2/2 + \uimm (\omega_{21} - \omega_{\mathrm{L}})]\bar{\tau}},\,\eu^{-[\gamma_3/2 + \uimm (\omega_{31} - \omega_{\mathrm{L}})]\bar{\tau}}\right)$. 

{In order to show how coherently controlled dynamics can be interpreted in terms of experimentally reconstructed SFI operators,} in Fig.~\ref{fig:twopulses} we present results for a sequence of two strong pulses aiming at the desired state $|\psi_{\mathrm{d}}\rangle = \sum_{i=1}^3 c_{\mathrm{d},i}|i\rangle$, of amplitudes $(c_{\mathrm{d},1}, \,c_{\mathrm{d},2}, \,c_{\mathrm{d},3}) = A\eu^{\uimm \gamma}\,(0, \,\sqrt{2/3}, \,\eu^{\uimm\delta}\sqrt{1/3})$, such that the ground state is completely depopulated, while the excited state $|2\rangle$ is twice as much populated as $|3\rangle$, despite a less favorable coupling to the ground state. The total final population $A^2$ and the phases $\gamma$ and $\delta$ are free parameters. Optimal pulse properties are determined via minimization of the cost function \cite{1367-2630-12-7-075008}
\begin{equation}
g(I_1, I_2, \bar{\tau}, \bar{\phi}) 
=\sqrt{ {\textstyle \sum_{i = 1}^3} \bigl| |{c}_{\mathrm{d},i}|^2 - |{c}_{\mathrm{p},i}|^2 \bigr|^2},
\label{eq:g}
\end{equation}
calculated for a discrete set of parameters and ensuring that $|\langle\psi_{\mathrm{d}} |\psi_{\mathrm{d}}\rangle|^2 = |\langle\psi_{\mathrm{p}} |\psi_{\mathrm{p}}\rangle|^2 = A^2$. 
A section of the control landscape \cite{1367-2630-12-7-075008}, associated with global minima of the cost function $g$, is displayed in Fig.~\ref{fig:twopulses}(b), confirming that it is a smooth function of its parameters, and small uncertainties in the pulse intensities do not lead to final states significantly differing from those expected. Figure~\ref{fig:twopulses}(c) shows the resulting dynamics of the system, when excited with the sequence of pulses determined via minimization of $g$, exhibiting very good agreement with the desired final state. {The displayed dynamics could be directly inferred from the reconstructed SFI operators. The state reached after the first intense pulse in Fig.~\ref{fig:twopulses}(c) is completely encoded in the matrix elements plotted in Fig.~\ref{fig:twopulses}(a), such that deterministic strong-field control provides an experiment-based visualization of the building blocks exploited by optimal control to reach a desired state. Rabi oscillations induced by strong ultrashort pulses are apparent in Fig.~\ref{fig:twopulses}(c), but knowledge of their explicit time dependence is not necessary to control the reached final state.} Although we focus on the control of final populations, this nevertheless requires phase knowledge of $\hat{U}^{\mathrm{R}}(I)$, such that $\hat{\varPhi}(\bar{\phi})$ and $\hat{V}(\bar{\tau})$ can ensure the necessary relative phase upon arrival of the second pulse \cite{tannor2007introduction}.

\begin{figure}[t]
\includegraphics[clip=true, width=\columnwidth]{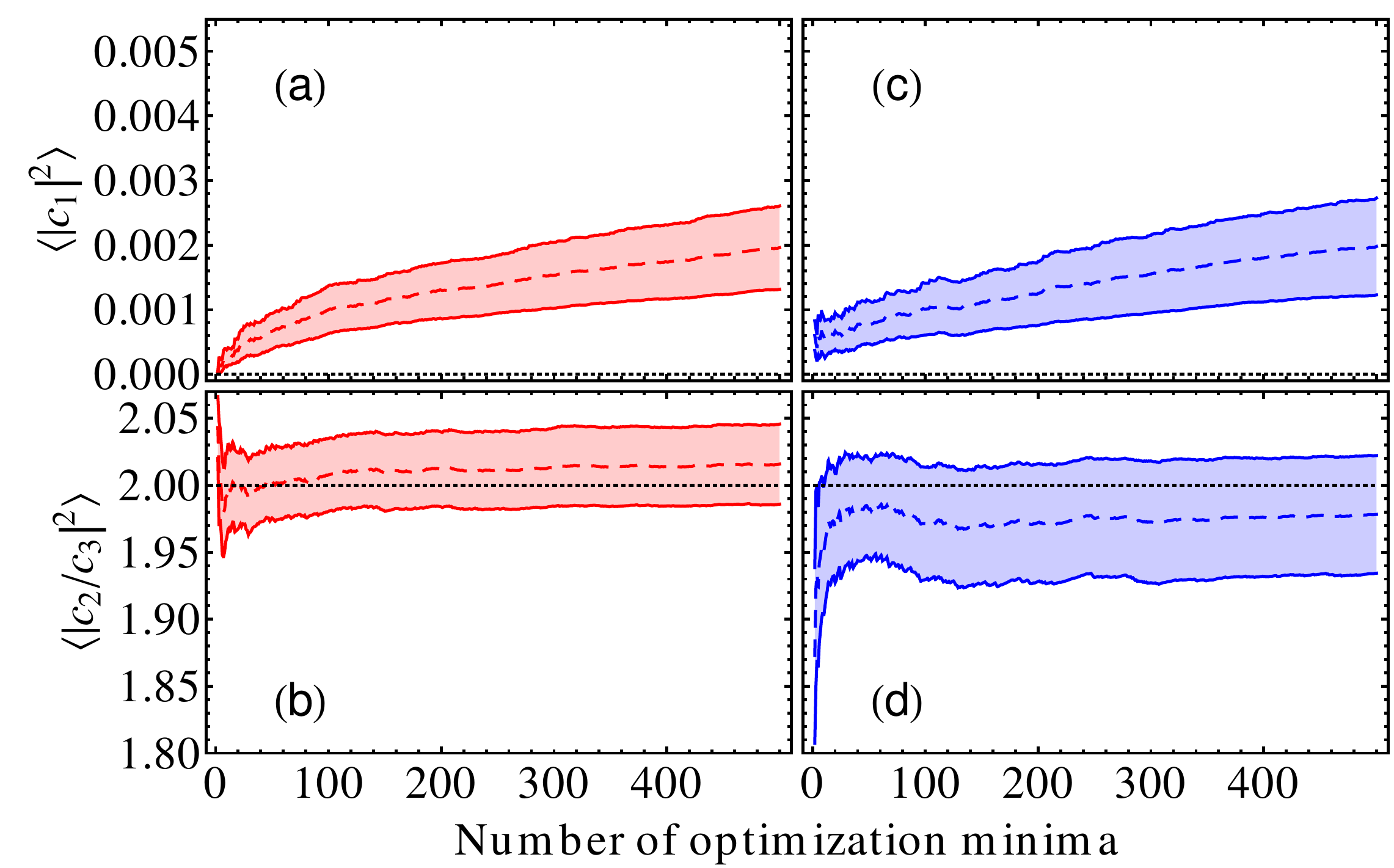}
\caption{(Color online). [(a),(c)] Reached ground-state population and [(b),(d)] ratio of the populations of the two excited states $|2\rangle$ and $|3\rangle$, averaged over the $N$ best sets of optimization pulse parameters $\{I_1,\,I_2,\,\bar{\tau},\,\bar{\phi}\}$, as a function of $N$. Mean values are displayed as dashed lines, the surrounding regions (bounded by continuous lines) have an amplitude given by the corresponding standard deviation. Desired final population and ratios are exhibited by black, dotted lines. Reached final populations and ratios are calculated for [(a),(b)] a three-level-only system and [(c),(d)] a five-level system. The best optimization parameters are obtained via minimization of the cost function~(\ref{eq:g}), calculated with SFI operators $\hat{U}^{\mathrm{R}}(I)$ reconstructed from transient-absorption spectra numerically simulated for [(a),(b)] a three-level and [(c),(d)] a five-level model.}
\label{fig:performances}
\end{figure} 

To verify the precision of the scheme, in Fig.~\ref{fig:performances}(a) and \ref{fig:performances}(b) we display final populations reached by the three-level system when using sequences of pulses determined through the minimization of Eq.~(\ref{eq:g}). The extracted SFI operators have an intrinsic uncertainty, and we therefore display results averaged among the first $N$ best sets of optimization parameters---with standard deviation---, as a function of $N$. Very good control performances are exhibited: complete depopulation of the ground state is reached [Fig.~\ref{fig:performances}(a)], and the mean value of the ratio $|c_2/c_3|^2$ is equal to 2 for the first best sets of optimization parameters, with relative uncertainty of $\sim 3\%$ [Fig.~\ref{fig:performances}(b)].

Finally, we test {our scheme in a realistic scenario, characterized by} the presence of incomplete modeling or perturbations. To derive $\mathcal{S}_{\mathrm{fit}}(\omega, \tau, U_{\mathrm{pu},ij})$ from Eq.~(\ref{eq:spectrum}), basic knowledge of the atomic transitions responsible for the absorption lines appearing in the spectrum is necessary. A robust control scheme should enable the manipulation of the states of interest also when additional, moderately contributing levels are present, which may not be known or experimentally discernible. As an example, we employ the fitting model $\mathcal{S}_{\mathrm{fit}}(\omega, \tau, U_{\mathrm{pu},ij})$ to extract $3\times3$ SFI operators $\hat{U}^{\mathrm{R}}$ from transient-absorption spectra in Rb atoms, stemming from the complete numerical simulation of the dynamics of the five-level system displayed in Fig.~\ref{fig:model}(b). 
The $E1$-allowed transitions $|2\rangle\rightarrow |4\rangle$, $|3\rangle\rightarrow |4\rangle$, and $|3\rangle\rightarrow |5\rangle$, with $\Delta M = 0$, with transition energies $\omega_{42} = 1.63\,\mathrm{eV}$ and $\omega_{53} = 1.60\,\mathrm{eV}$, are resonantly excited by the optical pulses, albeit more weakly than the $|1\rangle\rightarrow |k\rangle$ transitions, $k\in\{2,\,3\}$, owing to smaller dipole-moment matrix elements $\boldsymbol{D}_{kl} = D_{kl}\hat{\boldsymbol{e}}_z$, $l\in\{4,\,5\}$ \cite{PhysRevA.61.050502, PhysRevA.69.022509}. To ensure that this resonant coupling contributes moderately, we assume large linewidths $\gamma_{4}$ and $\gamma_5$, here set equal to $1/(100\,\mathrm{fs})$ \footnote{For instance, since the ionization potential of states \unexpanded{$|4\rangle$} and \unexpanded{$|5\rangle$} is $0.98\,\mathrm{eV}$, a laser tuned to (or slightly above) this energy would decrease population and coherence of these two excited states, without effectively affecting the remaining transitions at energies larger than $1.56\,\mathrm{eV}$.}, such that only the two lines associated with the $|1\rangle\rightarrow |k \rangle$ transitions can be clearly distinguished in the absorption spectra. Photoionization in the presence of an optical pulse is also accounted for \cite{LANL}. SFI operators $\hat{U}^{\mathrm{R}}(I)$ are extracted from these numerically calculated spectra, and used to control the three-level system in the box of Fig.~\ref{fig:model}(b) via minimization of the cost function $g$. The very good performances displayed in Figs.~\ref{fig:performances}(c) and \ref{fig:performances}(d) confirm that the method is robust and only marginally influenced by additional levels not accounted for explicitly in $\mathcal{S}_{\mathrm{fit}}(\omega,\tau,U_{\mathrm{pu},ij})$. Furthermore, in contrast to methods based exclusively on theory, maximal information on the strong-field interaction is extracted from the experimental spectra, including the background effect of unknown additional levels on the SFI operators $\hat{U}^{\mathrm{R}}(I)$ of interest. 

In conclusion, we have designed an optimized sequence for quantum-state control based on intensity-dependent operators extractable from observable transient-absorption spectra. Schemes consisting of a higher number of pulses are possible to further enhance the control precision or to achieve additional control goals simultaneously. The method was mainly discussed for a three-level scheme modeling Rb atoms, but this could be straightforwardly generalized to higher numbers of states. Our results are expected to trigger the development of related techniques for interaction-operator reconstruction of more complex systems such as molecules, for which strong-field absorption-line-shape control was recently demonstrated \cite{Meyer22122015}. The advances in coherent x-ray sources open up interesting prospects {especially} for the application of our method at short wavelengths. Quantifying the effect of strong broadband pulses from experimentally accessible spectra would then enable quantum control based on designed sequences of the available, ultrashort x-ray pulses, with added benefits such as site specificity near core transitions.

\begin{acknowledgments}
S.~M.~C. and Z.~H. acknowledge helpful discussions with J\"org Evers and Christian Ott. 
\end{acknowledgments}

\end{document}


\title{Supplemental Information to the article ``Deterministic strong-field quantum control''}

\author{Stefano M. Cavaletto}
\affiliation{Max-Planck-Institut f\"{u}r Kernphysik, Saupfercheckweg~1, 69117~Heidelberg, Germany}
\author{Zolt\'an Harman}
\affiliation{Max-Planck-Institut f\"{u}r Kernphysik, Saupfercheckweg~1, 69117~Heidelberg, Germany}
\author{Thomas Pfeifer}
\affiliation{Max-Planck-Institut f\"{u}r Kernphysik, Saupfercheckweg~1, 69117~Heidelberg, Germany}
\author{Christoph H. Keitel}
\affiliation{Max-Planck-Institut f\"{u}r Kernphysik, Saupfercheckweg~1, 69117~Heidelberg, Germany}


\maketitle


In Sec.~\ref{Equations of motion and transient-absorption spectra}, details on the numerical calculation of transient-absorption spectra $\mathcal{S}(\omega, \tau)$ are given, which we use to simulate experimental absorption spectra and, thus, validate our operator-reconstruction method. In Sec.~\ref{Fitting analytical model of transient-absorption spectra in terms of intensity-dependent control operators}, an analytical fitting model $\mathcal{S}_{\mathrm{fit}}(\omega, \tau, U_{\mathrm{pu},ij})$ is provided, which we use to fit data from transient-absorption spectroscopy (TAS) and reconstruct interaction operators $\hat{U}(I)$. Finally, in Sec.~\ref{Density-matrix formulation}, we generalize our formalism to cases in which a density-matrix approach is required.

\section{Numerical simulations of the quantum dynamics and the associated transient-absorption spectra}
\label{Equations of motion and transient-absorption spectra}

\setcounter{equation}{0}

We model Rb atoms as three- or five-level systems, as shown in Fig.~2(b) in the article. We consider a quantum system in the state $|\psi(t)\rangle = \hat{\mathcal{U}}(t,t_0)|\psi(t_0)\rangle$, where the evolution operator $\hat{\mathcal{U}}(t,t_0)$, describing the dynamics from $t_0$ to $t$, is solution of the Schr\"odinger equation 
\begin{equation}
\frac{\diff \hat{\mathcal{U}}(t,t_0)}{ \diff t}  = -\uimm[\hat{H}_0 + \hat{H}_{\mathrm{int}}]\,\hat{\mathcal{U}}(t,t_0)\ \ \ \hat{\mathcal{U}}(t_0,t_0) = \hat{I}.
\label{eq:evolution-operator}
\end{equation}
The total Hamiltonian consists of the atomic-structure Hamiltonian $\hat{H}_0 = \sum_i{(\omega_{i} -\uimm\gamma_i/2 ) |i \rangle\langle i|}$, complex-valued and including energies $\omega_{i}$ and linewidths $\gamma_{i}$ of each atomic state $|i\rangle$, and of the electric-dipole ($E1$) light-matter interaction $\hat{H}_{\mathrm{int}} = - [\hat{\boldsymbol{D}}^{-}\cdot\,\boldsymbol{\mathcal{E}}^{+}(t) + \mathrm{H.\,c.}]$, in the rotating-wave approximation \cite{Scully:QuantumOptics}. Here, $\hat{I} = \sum_i{|i \rangle\langle i|}$ is the identity operator, while $\hat{\boldsymbol{D}}^{-} = \sum_{i>j} \boldsymbol{D}_{ij} |i \rangle\langle j|$ is the negative-frequency part of the dipole-moment operator, of matrix elements $\boldsymbol{D}_{ij}$, and $\boldsymbol{\mathcal{E}}^{+}(t)$ is the complex electric field \cite{diels2006ultrashort}. Pump and probe pulses are modeled by functions of the form $\boldsymbol{\mathcal{E}}(t-t_{\mathrm{c}},\phi) = \boldsymbol{\mathcal{E}}^{+}(t -t_{\mathrm{c}},\phi ) + \mathrm{c.c.} = \mathcal{E}_0\,f\left(t-t_{\mathrm{c}}\right)\,\cos\left[\omega_{\mathrm{L}} \left(t-t_{\mathrm{c}}\right)+\phi\right]\hat{\boldsymbol{e}}_z$, centered around $t_{\mathrm{c}}$, with carrier-envelope phase (CEP) $\phi$ and envelope function of the form $f(t) = \cos^2{(\pi t/T)}\,R(t/T)$. Here, $R(x) = \theta(x+1/2)-\theta(x-1/2)$ is defined in terms of the Heaviside step function $\theta(x)$, such that $T = \pi T_{\mathrm{FWHM}}/[2\arccos{(\sqrt[4]{1/2})}]$, with $T_{\mathrm{FWHM}}$ being the full width at half maximum of $f^2(t)$. The evolution operator $\hat{\mathcal{U}}_{0}(t, t_0)$ introduced in the article is associated with a pulse centered on $t_{\mathrm{c}} = 0$ and with vanishing CEP $\phi = 0$. The generic evolution operator reads $\hat{\mathcal{U}}(t, t_0) = \hat{\varPhi}\daga(\phi)\,\hat{\mathcal{U}}_0\left(t - t_{\mathrm{c}},t_0-t_{\mathrm{c}}\right)\,\hat{\varPhi}(\phi)$, with the diagonal operator $\hat{\varPhi}(\phi)$ accounting for the pulse CEP.

{In order to better present our key idea, we employ a Schr\"odinger-equation-based formalism throughout the manuscript, with decay rates $\gamma_2 = \gamma_3 = 1/(1500\,\mathrm{fs})$ modeling experimental linewidth due to, e.g., Doppler or collision-induced broadening. Radiative decay times of $27\,\mathrm{ns}$ and $25.7\,\mathrm{ns}$, from the excited states 2 and 3, respectively, are more than 4 orders of magnitude larger than the duration of the femtosecond pulses used for our control scheme. Therefore, spontaneous decay can be neglected for our system. In this time scale, also the employed decay rates $\gamma_2 = \gamma_3 = 1/(1500\,\mathrm{fs})$ are small and the difference with a density-matrix treatment is therefore negligible. In Sec.~\ref{Density-matrix formulation} we display how our results can be generalized to a more complex case requiring a density-matrix formulation.}

We solve the Schr\"odinger equation~(\ref{eq:evolution-operator}) to calculate the time evolution of the system interacting with delayed pump and probe pulses, and use the obtained solution to simulate transient-absorption spectra. The same procedure is used to calculate the dynamics of the controlled quantum system, displayed in Figs.~3(c) and 4 in the article, for an optimal sequence of two intense pulses.

To avoid repetitions, in the following we immediately refer to the more general case of a five-level scheme.  
The state of the system $|\psi^{(5)}(t,\tau)\rangle = \sum_{i=1}^5c^{(5)}_i(t,\tau)|i\rangle$ has a vector representation given by $\boldsymbol{c}^{(5)} = (c^{(5)}_1,\,c^{(5)}_2,\,c^{(5)}_3,\,c^{(5)}_4,\,c^{(5)}_5)^{\mathrm{T}}$. The free-evolution operator is given by 
\begin{widetext}
\begin{equation}
\hat{V}^{(5)}(t) = \diag (1,\,\eu^{-(\frac{\gamma_2}{2} + \uimm \omega_{21})t},\,\eu^{-(\frac{\gamma_3}{2} + \uimm \omega_{31})t},\,\eu^{-(\frac{\gamma_4}{2} + \uimm \omega_{41})t},\,\eu^{-(\frac{\gamma_5}{2} + \uimm \omega_{51})t}).
\end{equation}
Here, the two highest-lying excited states are almost degenerate, with transition energies $\omega_{41} = \omega_{51} = 3.19\,\mathrm{eV}$. For the $E1$-allowed transitions depicted in Fig.~2(b) in the article, we introduce the time-dependent Rabi frequencies $\varOmega_{ij}(t) = D_{ij}\,\mathcal{E}_0\,f(t)$, with dipole-moment matrix elements $\hat{\boldsymbol{D}}_{ij} = D_{ij}\hat{\boldsymbol{e}}_z$ aligned along the pulse polarization vector \cite{PhysRevA.61.050502, PhysRevA.69.022509}. We calculate the time evolution of the atomic system interacting with a pump pulse centered on $t_{\mathrm{c}} = 0$ and a probe pulse centered on $t_{\mathrm{c}} =\tau$. We assume a weak probe pulse, with duration $T_{\mathrm{FWHM, pr}} = 15\,\mathrm{fs}$ and intensity $I_{\mathrm{pr}} = 1\times10^8\,\mathrm{W/cm^2}$, and pump pulses of duration $T_{\mathrm{FWHM, pu}} = 30\,\mathrm{fs}$ and intensities $I_{\mathrm{pu}}$ varying between $0.1\times10^{10}\,\mathrm{W/cm^2}$ and $5\times10^{10}\,\mathrm{W/cm^2}$. For $(t-t_{\mathrm{c}})\in[-T/2,\,T/2]$, the time evolution of $\boldsymbol{c}^{(5)}$ in the rotating-wave approximation is given by 
\begin{equation}
\frac{\diff \boldsymbol{c}^{(5)}}{\diff t} = [\hat{\varPhi}^{(5)}(\phi)]\daga\,[\hat{\varLambda}^{(5)}(t-t_{\mathrm{c}})]\daga\,\hat{M}^{(5)}(t-t_{\mathrm{c}})\,\hat{\varLambda}^{(5)}(t-t_{\mathrm{c}})\,\hat{\varPhi}^{(5)}(\phi )\,\boldsymbol{c}^{(5)}(t),
\label{eq:timeevolution5}
\end{equation}
with the phase of the pulse accounted for by $\hat{\varPhi}^{(5)}(\phi) = \diag(1,\,\eu^{\uimm\phi},\,\eu^{\uimm\phi},\,\eu^{2\uimm\phi},\,\eu^{2\uimm\phi})$ and $\hat{\varLambda}^{(5)}(t) = \diag(1,\,\eu^{\uimm\omega_{\mathrm{L}}t},\,\eu^{\uimm\omega_{\mathrm{L}}t},\,\eu^{2\uimm\omega_{\mathrm{L}}t},\,\eu^{2\uimm\omega_{\mathrm{L}}t})$, while
\begin{equation}
 \hat{M}^{(5)}(t) = 
\begin{pmatrix}
  0 & \uimm \frac{\varOmega_{12}(t)}{2} & \uimm \frac{\varOmega_{13}(t)}{2} & 0 & 0\\
  \uimm \frac{\varOmega_{12}^*(t)}{2} &-\frac{\gamma_2}{2}-\uimm \omega_{21} & 0& \uimm \frac{\varOmega_{24}(t)}{2} & 0\\
  \uimm \frac{\varOmega_{13}^*(t)}{2} & 0 &-\frac{\gamma_3}{2}-\uimm \omega_{31} & \uimm \frac{\varOmega_{34}(t)}{2} & \uimm \frac{\varOmega_{35}(t)}{2}\\
  0 & \uimm \frac{\varOmega_{24}^*(t)}{2} &\uimm \frac{\varOmega_{34}^*(t)}{2} & -\frac{\gamma_4 + \gamma_{\mathrm{P},4}(t)}{2}-\uimm \omega_{41} & 0\\
  0 & 0 &\uimm \frac{\varOmega_{35}^*(t)}{2} & 0 & -\frac{\gamma_5 + \gamma_{\mathrm{P},5}(t)}{2}-\uimm \omega_{51}
\end{pmatrix}.
\end{equation}
\end{widetext}
Photoionization of states $|4\rangle$ and $|5\rangle$ in the presence of an optical pulse has also been included as an effective loss of amplitude at the rate $\gamma_{\mathrm{P},l}(t) = \sigma_l(\omega_{L}) \mathcal{I}_{\mathrm{L}}(t)$, with photoionization cross sections $\sigma_l(\omega_{L})$ calculated with \cite{LANL}, and optical flux $\mathcal{I}_{\mathrm{L}}(t) = I_{\mathrm{L}}(t)/\omega_{L}$ defined in terms of the time-dependent pulse intensity $I_{L}(t) = [\mathcal{E}_0\,f(t)]^2/(8\pi\alpha)$. 

The dynamics of the five-level system are associated with the $5\times 5$ evolution operator $\hat{\mathcal{U}}^{(5)}(t, t_0)$ and the related interaction operator 
\begin{equation}
\begin{split}
\hat{U}^{(5)} &= \hat{V}^{(5)}(-T/2)\,\hat{\mathcal{U}}^{(5)}_{0}(T/2, -T/2)\,[\hat{V}^{(5)}(T/2)]^{-1} \\
&=
\begin{pmatrix}
   \hat{U}^{(5)}_{3\times3} & \hat{U}^{(5)}_{3\times2}\\
   \hat{U}^{(5)}_{2\times3} & \hat{U}^{(5)}_{2\times2}                        
\end{pmatrix}.
\end{split}
\end{equation}
The time evolution of the five-level system, calculated from Eq.~(\ref{eq:timeevolution5}), is used to numerically simulate transient-absorption spectra $\mathcal{S}^{(5)}(\omega,\tau)$. 
Fast oscillations as a function of the time delay $\tau$ are averaged out in a noncollinear geometry \cite{PhysRevLett.115.033003} and cannot be discerned. We numerically calculate the averaged time-delay-dependent spectrum by performing a convolution of the fast oscillating spectrum with a normalized Gaussian function $G(\tau, \Delta\tau)$ of width $\Delta\tau = 5\times2\pi/\omega_{\mathrm{L}}$ \cite{PhysRevLett.115.033003}. 


For a three-level model associated with the state $\left|\psi^{(3)}(t,\tau)\right\rangle =\sum_{i=1}^3 c^{(3)}_i(t,\tau)|i\rangle$ and the density-matrix elements $\rho^{(3)}_{ij}(t,\tau)= c^{(3)}_i(t,\tau)\,[c^{(3)}_j(t,\tau)]^*$, $i,\,j\in\{1,\,2,\,3\}$, then the absorption spectrum results from
\begin{equation}
\mathcal{S}^{(3)}(\omega,\tau)\propto -\omega\Imm\left[\sum_{k=2}^3 D_{1k}^* \int_{-\infty}^{\infty}\rho^{(3)}_{1k}(t,\tau)\,\eu^{-\uimm \omega (t- \tau)}\diff t\right].
\label{eq:spectrum3}
\end{equation}
Similarly, the dynamics of the three-level system is described in terms of the operators $\hat{V}^{(3)}(t) = \diag (1,\,\eu^{-(\frac{\gamma_2}{2} + \uimm \omega_{21})t},\,\eu^{-(\frac{\gamma_3}{2} + \uimm \omega_{31})t})$, $\hat{\varPhi}^{(3)}(\phi) = \diag(1,\,\eu^{\uimm\phi},\,\eu^{\uimm\phi})$,  $\hat{\varLambda}^{(3)}(t) = \diag(1,\,\eu^{\uimm\omega_{\mathrm{L}}t},\,\eu^{\uimm\omega_{\mathrm{L}}t})$, while
\begin{equation}
 \hat{M}^{(3)}(t) = 
\begin{pmatrix}
  0 & \uimm \frac{\varOmega_{12}(t)}{2} & \uimm \frac{\varOmega_{13}(t)}{2} \\
  \uimm \frac{\varOmega_{12}^*(t)}{2} &-\frac{\gamma_2}{2}-\uimm \omega_{21} & 0\\
  \uimm \frac{\varOmega_{13}^*(t)}{2} & 0 &-\frac{\gamma_3}{2}-\uimm \omega_{31}
\end{pmatrix}, 
\label{eq:M3}
\end{equation}
resulting in a $3\times 3$ evolution operator $\hat{\mathcal{U}}^{(3)}(t, t_0)$ and a related interaction operator 
$\hat{U}^{(3)} = \hat{V}^{(3)}(-T/2)\,\hat{\mathcal{U}}^{(3)}_{0}(T/2, -T/2)\,[\hat{V}^{(3)}(T/2)]^{-1}$. 
It is important to stress that the $3\times 3$ matrix $\hat{U}^{(5)}_{3\times3}$ is different from $\hat{U}^{(3)}$, since $\hat{U}^{(5)}_{3\times3}$ is influenced by the presence of states $|4\rangle$ and $|5\rangle$, which could not be taken into account if one solved the Schr\"odinger equation for a three-level system exclusively. By using TAS to extract strong-field interaction (SFI) operators by means of the analytical fitting model presented in Sec.~\ref{Fitting analytical model of transient-absorption spectra in terms of intensity-dependent control operators}, one has thus access to $\hat{U}^{(5)}_{3\times3}$, in contrast to methods based exclusively on a theory which could only provide $\hat{U}^{(3)}$. 

\section{Analytical fitting model of transient-absorption spectra in terms of interaction operators}
\label{Fitting analytical model of transient-absorption spectra in terms of intensity-dependent control operators}

To interpret transient-absorption spectra from Rb atoms, modeled by the few-level scheme depicted in Fig.~2(b) in the article, we focus on a $V$-type three-level model, accounting for the two transitions $|1\rangle\rightarrow|k\rangle$, $k\in\{2,\,3\}$. 
Starting from these assumptions and the associated definition of the spectrum in Eq.~(\ref{eq:spectrum3}), we derive an analytical fitting model to relate experimental transient-absorption spectra to the operators $\hat{U}_{\mathrm{pu}}(I)$ and $\hat{U}_{\mathrm{pr}}$.

\subsection{Interaction operator for weak short probe pulses}
\label{Evolution operator for weak short probe pulses}

The operator $\hat{U}_{\mathrm{pr}}$ models a weak probe pulse, with FWHM of $15\,\mathrm{fs}$ and intensity of $1\times 10^{8}\,\mathrm{W/cm^2}$. In order to interpret results from TAS, we approximate the envelope function $f_{\mathrm{pr}}(t)$ of such a short (broadband) weak pulse with a Dirac-$\delta$-like peak
\begin{equation}
\tilde{f}_{\mathrm{pr}}(t) = \delta(t)\,\int_{-T_{\mathrm{pr}}/2}^{T_{\mathrm{pr}}/2} f_{\mathrm{pr}}(t')\,\diff t',
\end{equation}
where $T_{\mathrm{pr}}$ is the duration of the pulse. 

An explicit solution of Eq.~(\ref{eq:evolution-operator}) provides
\begin{equation}
\hat{\mathcal{U}}_{\mathrm{pr}}(t,t_0) = \hat{V}(t)\,\eu^{\hat{B}\,[\theta(t) - \theta(t_0 )]}\,\hat{V}(-t_0),
\end{equation}
where $\theta(x)$ is the Heaviside step function, $\hat{V}(t)$ is the free-evolution operator, and 
\begin{equation}
\hat{B} = 
\begin{pmatrix}
  0 & \uimm \frac{\vartheta_2}{2} & \uimm \frac{\vartheta_3}{2}\\
  \uimm \frac{\vartheta_2^*}{2} &0 & 0\\
  \uimm \frac{\vartheta_3^*}{2} & 0 &0
\end{pmatrix}
\end{equation}
is defined in terms of the pulse areas $\vartheta_k =\int_{-T_{\mathrm{pr}}/2}^{T_{\mathrm{pr}}/2} \varOmega_{\mathrm{pr},1k}(t)\,\diff t$ of the probe-pulse Rabi frequencies $\varOmega_{\mathrm{pr},1k}(t)$, for $k\in\{2,\,3\}$. Taking advantage of the weak probe-pulse intensity, the probe operator $\hat{U}_{\mathrm{pr}} = \hat{V}(-T_{\mathrm{pr}}/2)\,\hat{\mathcal{U}}_{\mathrm{pr}}(T_{\mathrm{pr}}/2, -T_{\mathrm{pr}}/2)\,\hat{V}^{-1}(T_{\mathrm{pr}}/2)$ reads
\begin{equation}
\hat{U}_{\mathrm{pr}} =  \eu^{\hat{B}} \approx \hat{I}_{3} + \hat{B} =
\begin{pmatrix}
  1 & \uimm \frac{\vartheta_2}{2} & \uimm \frac{\vartheta_3}{2}\\
  \uimm \frac{\vartheta_2^*}{2} &1 & 0\\
  \uimm \frac{\vartheta_3^*}{2} & 0 &1
\end{pmatrix}.
\label{eq:Upr}
\end{equation}

\subsection{Interpretation of transient-absorption spectra and reconstruction of strong-field interaction operators}
\label{Transient-absorption spectra}

While first-order perturbation theory is used to write $\hat{U}_{\mathrm{pr}}$ explicitly [Eq.~(\ref{eq:Upr})], no assumption is required for the operator $\hat{U}_{\mathrm{pu}}$, describing the interaction with strong pump pulses. We show how its 9 matrix elements are linked to measurable transient-absorption spectra, deriving an analytical fitting model which can be used to extract $\hat{U}_{\mathrm{pu}}$ directly from observable data.

\subsubsection{Probe-pump scheme}
For a probe-pump scheme ($\tau<0$), the weak probe pulse generates the initial excited state, which is nonperturbatively modified by the action of the second-arriving intense pump pulse. In terms of the operators $\hat{U}_{\mathrm{pr}}$ and $\hat{U}_{\mathrm{pu}}(I)$, modeling light-matter interactions as effectively instantaneous transformations, the effective evolution of the time-delay-dependent state $\left|\psi_{\mathrm{fit}}(t,\tau)\right\rangle = \sum_{i=1}^3 c_i(t,\tau)|i\rangle$ from the effective initial state $|\psi_0\rangle = |1\rangle$ is given by
\begin{equation}
|\psi_{\mathrm{fit}}(t,\tau)\rangle = \left\{
\begin{aligned}
&|\psi_0\rangle, &\text{if $t<\tau$},\\
&\hat{V}(t-\tau)\hat{U}_{\mathrm{pr}}|\psi_0\rangle, &\text{if $\tau<t<0$},\\
&\hat{V}(t)\hat{U}_{\mathrm{pu}}(I)\,\hat{V}(-\tau)\,\hat{U}_{\mathrm{pr}}|\psi_0\rangle, &\text{if $t>0$}.
\end{aligned}
\right.
\label{eq:evolutionprobepump}
\end{equation}
We can analogously define the effective density-matrix elements $\rho_{\mathrm{fit},ij}(t,\tau)= c_i(t,\tau)\,[c_j(t,\tau)]^*$, in terms of the components $c_{i}(t,\tau)$ of $\left|\psi_{\mathrm{fit}}(t,\tau)\right\rangle$. By inserting $\rho_{\mathrm{fit},ij}(t,\tau)$ into Eq.~(\ref{eq:spectrum3}), an analytical interpretation model for the probe-pump spectrum is obtained as
\begin{equation}
\begin{split}
&\mathcal{S}_{\mathrm{fit}}(\omega,\tau,U_{\mathrm{pu},ij}) \\
\propto\,& -\omega\Imm\biggl\{\sum_{k=2}^3 D_{1k}^* \biggr[\int_{\tau}^{0}\rho_{\mathrm{fit},1k}(t,\tau)\,\eu^{-\uimm\omega (t-\tau)}\,\diff t \\
&\ \ \ \ \ \ \ \ \ \ \ \ + \int_{0}^{\infty}\rho_{\mathrm{fit},1k}(t,\tau)\,\eu^{-\uimm\omega (t-\tau)}\,\diff t\biggr]\biggr\}\,.
\end{split}
\label{eq:spprpu}
\end{equation}
The first integral in Eq.~(\ref{eq:spprpu}) is equal to
\begin{equation}
\begin{split}
&\,\int_{\tau}^{0}\rho_{\mathrm{fit},1k}(t,\tau)\,\eu^{-\uimm\omega (t-\tau)}\,\diff t \\
=&\,U_{\mathrm{pr,}11}\, U^*_{\mathrm{pr,}k1} \int_{\tau}^{0} V^*_{kk}(t-\tau) \,\eu^{-\uimm\omega (t-\tau)}\,\diff t \\
= &\, -\uimm\frac{\vartheta_k}{2}\,\frac{1-\eu^{\uimm(\omega - \omega_{k1}) \tau}\,\eu^{\frac{\gamma_k}{2}\tau}}{\uimm(\omega - \omega_{k1}) + \frac{\gamma_k}{2}},
\end{split}
\label{eq:spprpu1stcomponent}
\end{equation}
which, for $\omega \approx \omega_{k1}$, does not feature fast oscillations as a function of the time delay $\tau$. The second integral in Eq.~(\ref{eq:spprpu}) is given by
\begin{equation}
\begin{split}
&\,\int_{0}^{\infty}\rho_{\mathrm{fit},1k}(t,\tau)\,\eu^{-\uimm\omega (t-\tau)}\,\diff t \\
=&\,\sum_{j,j' =1}^3U_{\mathrm{pu},1j}\,U_{\mathrm{pu},kj'}^*\,V_{jj}(-\tau)\,V_{j'j'}^*(-\tau)\,U_{\mathrm{pr},j1}\,U_{\mathrm{pr},j'1}^*\\
&\times\,\int_{0}^{\infty} V_{kk}^*(t)\,\eu^{-\uimm\omega (t-\tau)} \,\diff t \\
= &\ \frac{1}{\,\uimm(\omega - \omega_{k1}) + \frac{\gamma_k}{2}}\sum_{j,j' =1}^3U_{\mathrm{pu},1j}\,U_{\mathrm{pu},kj'}^*\,\\
&\ \ \times\,\left[V_{jj}(-\tau)\,V_{j'j'}^*(-\tau)\,\eu^{\uimm\omega\tau}\right]\,U_{\mathrm{pr},j1}\,U_{\mathrm{pr},j'1}^*. 
\end{split}
\label{eq:spprpu2ndcomponent}
\end{equation}
Firstly, all terms in the above sum depend upon nonvanishing elements of the probe-pulse operator $\hat{U}_{\mathrm{pr}}$ [Eq.~(\ref{eq:Upr})]. In a collinear geometry all terms in the above sum contribute to the resulting transient-absorption spectrum, which can be used to extract SFI matrix-element products such as $U_{\mathrm{pu}, 1j}\,U_{\mathrm{pu}, kj'}$. In a noncollinear geometry, however, such as the one utilized in Ref.~\cite{PhysRevLett.115.033003}, fast oscillating terms due to $[V_{jj}(-\tau)\,V_{j'j'}^*(-\tau)\,\eu^{\uimm\omega\tau}]$ are averaged out and do not contribute to the resulting spectra. Equation~(\ref{eq:spprpu2ndcomponent}) can be employed to recognize and eliminate all terms which, at $\omega\approx \omega_{k1}$, feature fast oscillations in $\tau$. As a result, 
in terms of the two functions
\begin{equation}
\begin{split}
A_2(\tau) &= U_{\mathrm{pu,}11}\,U^*_{\mathrm{pu,}22} + U_{\mathrm{pu,}11}\,U^*_{\mathrm{pu,}23}\, \frac{\vartheta_{3}}{\vartheta_{2}}\, \eu^{-\uimm\omega_{32}\tau}\,\eu^{\frac{\gamma_3-\gamma_2}{2}\tau} ,\\
A_3(\tau) &= U_{\mathrm{pu,}11}\,U^*_{\mathrm{pu,}33}+ U_{\mathrm{pu,}11}\,U^*_{\mathrm{pu,}32}\,\frac{\vartheta_{2}}{\vartheta_{3}}\, \eu^{\uimm\omega_{32}\tau}\,\eu^{\frac{\gamma_2-\gamma_3}{2}\tau} ,
\end{split}
\end{equation}
average experimental spectra can be modeled by
\begin{equation}
\begin{split}
\langle \mathcal{S}_{\mathrm{fit}}(\omega,\tau,U_{\mathrm{pu},ij})\rangle_{\tau} & \propto-\omega\Imm\biggl\{\sum_{k=2}^3\frac{-\uimm\,D_{1k}^*\frac{\vartheta_k}{2}}{\uimm(\omega - \omega_{k1}) + \frac{\gamma_k}{2}}\\
&\ \times \bigl[1+\eu^{\uimm(\omega - \omega_{k1})\tau}\,\eu^{\frac{\gamma_k}{2}\tau}\, (A_k(\tau)-1)\bigr]\biggr\}.
\end{split}
\label{eq:Sexpprpu}
\end{equation}
The proportionality symbol $\propto$ stresses that these spectra depend on a density-dependent multiplication factor $K$, which we treat here as a fitting parameter, along with the SFI matrix-element products $U_{\mathrm{pu,}11}\,U^*_{\mathrm{pu,}22}$, $U_{\mathrm{pu,}11}\,U^*_{\mathrm{pu,}33}$, $U_{\mathrm{pu,}11}\,U^*_{\mathrm{pu,}32}$, and $U_{\mathrm{pu,}11}\,U^*_{\mathrm{pu,}23}$.

\subsubsection{Pump-probe scheme}

For a pump-probe scheme ($\tau>0$), in terms of the operators $\hat{U}_{\mathrm{pr}}$ and $\hat{U}_{\mathrm{pu}}(I)$, the effective evolution of the time-delay-dependent state $\left|\psi_{\mathrm{fit}}(t,\tau)\right\rangle $ from the effective initial state $|\psi_0\rangle = |1\rangle$ can be modeled as
\begin{equation}
|\psi_{\mathrm{fit}}(t,\tau)\rangle = \left\{
\begin{aligned}
&|\psi_0\rangle, &\text{if $t<0$},\\
&\hat{V}(t)\hat{U}_{\mathrm{pu}}(I)|\psi_0\rangle, &\text{if $0<t<\tau$},\\
&\hat{V}(t-\tau)\hat{U}_{\mathrm{pr}}\,\hat{V}(\tau)\,\hat{U}_{\mathrm{pu}}(I)|\psi_0\rangle, &\text{if $t>\tau$}.
\end{aligned}
\right.
\label{eq:evolutionpumpprobe}
\end{equation}
By including in Eq.~(\ref{eq:spectrum3}) the effective evolution of the matrix elements $\rho_{\mathrm{fit},ij}(t,\tau)$ from Eq.~(\ref{eq:evolutionpumpprobe}), an analytical interpretation model for the pump-probe spectrum is derived, which can be split into the following sum:
\begin{equation}
\begin{split}
&\mathcal{S}_{\mathrm{fit}}(\omega,\tau, U_{\mathrm{pu},ij}) \\
\propto\,& -\omega\Imm\biggl\{\sum_{k=2}^3 D_{1k}^* \biggr[\int_{0}^{\tau}\rho_{\mathrm{fit},1k}(t,\tau)\,\eu^{-\uimm\omega (t-\tau)}\,\diff t \\
&\ \ \ \ \ \ \ \ \ + \int_{\tau}^{\infty}\rho_{\mathrm{fit},1k}(t,\tau)\,\eu^{-\uimm\omega (t-\tau)}\,\diff t\biggr]\biggr\}\,.
\end{split}
\label{eq:sppupr}
\end{equation}
The first integral in Eq.~(\ref{eq:sppupr}) is equal to
\begin{equation}
\begin{split}
&\,\int_{0}^{\tau}\rho_{\mathrm{fit},1k}(t,\tau)\,\eu^{-\uimm\omega (t-\tau)}\,\diff t \\
=&\,U_{\mathrm{pu,}11}\,U^*_{\mathrm{pu,}k1}\,\int_{0}^{\tau}V_{kk}^*(t)\,\eu^{-\uimm\omega (t-\tau)}\,\diff t \\
= &\, U_{\mathrm{pu,}11}\,U^*_{\mathrm{pu,}k1}\,\eu^{\uimm\omega \tau}\,\frac{1-\eu^{-\uimm(\omega - \omega_{k1}) \tau}\,\eu^{-\frac{\gamma_k}{2}\tau}}{\uimm(\omega - \omega_{k1}) + \frac{\gamma_k}{2}}.
\end{split}
\label{eq:sppupr1stcomponent}
\end{equation}
As described in the case of the probe-pump scheme, the fast oscillations at frequencies $\omega \approx \omega_{k1}$ are averaged out in a noncollinear geometry, and this first integral does not contribute to the associated average absorption spectrum. The second integral in Eq.~(\ref{eq:sppupr}) is given by
\begin{equation}
\begin{split}
&\,\int_{\tau}^{\infty}\rho_{\mathrm{fit},1k}(t,\tau)\,\eu^{-\uimm\omega (t-\tau)}\,\diff t \\
= &\,\sum_{j,j' =1}^3U_{\mathrm{pr},1j}\,U_{\mathrm{pr},kj'}^*\,V_{jj}(\tau)\,V_{j'j'}^*(\tau)\,U_{\mathrm{pu},j1}\,U_{\mathrm{pu},j'1}^*\\
&\times\,\int_{\tau}^{\infty} V_{kk}^*(t-\tau)\,\eu^{-\uimm\omega (t-\tau)} \,\diff t \\
= &\,\frac{1}{\,\uimm(\omega - \omega_{k1}) + \frac{\gamma_k}{2}} \sum_{j,j' =1}^3U_{\mathrm{pr},1j}\,U_{\mathrm{pr},kj'}^*\,\\
&\times\,\left[V_{jj}(\tau)\,V_{j'j'}^*(\tau)\right]\,U_{\mathrm{pu},j1}\,U_{\mathrm{pu},j'1}^*. 
\end{split}
\label{eq:sppupr2ndcomponent}
\end{equation}
By using Eq.~(\ref{eq:Upr}) and removing fast oscillating terms due to $V_{jj}(\tau)\,V_{j'j'}^*(\tau)$, which would not appear in a noncollinear geometry, we conclude that
\begin{equation}
\begin{split}
&\langle \mathcal{S}_{\mathrm{fit}}(\omega,\tau, U_{\mathrm{pu},ij})\rangle_{\tau} \\
\propto &-\omega\Imm\biggl\{\frac{D_{12}^*}{\uimm(\omega - \omega_{21}) + \frac{\gamma_2}{2}}\biggl[ \Bigl( -\uimm\frac{\vartheta_{2}}{2}\, |U_{\mathrm{pu,}11}|^2\\
&\ \ \ \ \ \ \ \ \  +\uimm\frac{\vartheta_{2}}{2}\, |U_{\mathrm{pu,}21}|^2 \,\eu^{-\gamma_2\tau} \Bigr)\\
&\ \ \ \ \ \ \ \ \  +\uimm\frac{\vartheta_{3}}{2}\,U_{\mathrm{pu,}31} \,U^*_{\mathrm{pu,}21} \,\eu^{-\uimm\omega_{32}\tau}\,\eu^{-\frac{\gamma_2+\gamma_3}{2}\tau} \biggr]\\
&\ \ \ \ \  + \frac{D_{13}^*}{\uimm(\omega - \omega_{31}) + \frac{\gamma_3}{2}}\biggl[ \Bigl( -\uimm\frac{\vartheta_{3}}{2}\, |U_{\mathrm{pu,}11}|^2\\
&\ \ \ \ \ \ \ \ \  +\uimm\frac{\vartheta_{3}}{2}\, |U_{\mathrm{pu,}31}|^2 \,\eu^{-\gamma_3\tau}\Bigr)\\
&\ \ \ \ \ \ \ \ \  +\uimm\frac{\vartheta_{2}}{2}\, U_{\mathrm{pu,}21}\,U^*_{\mathrm{pu,}31}\,\eu^{\uimm\omega_{32}\tau}\,\eu^{-\frac{\gamma_2+\gamma_3}{2}\tau}\biggr]\biggr\},
\end{split}
\label{eq:Sexppupr}
\end{equation}
which explicitly depends upon $|U_{\mathrm{pu,}11}|^2$, $|U_{\mathrm{pu,}22}|^2$, $|U_{\mathrm{pu,}33}|^2$, and $(U_{\mathrm{pu,}31} \,U^*_{\mathrm{pu,}21})$. The same multiplication factor $K$ should be used which was extracted from the probe-pump spectrum [see Eq.~(\ref{eq:Sexpprpu})].

\subsubsection{Additional remarks}

The matrix elements of the SFI operator $\hat{U}_{\mathrm{pu}}$ are reconstructed by using the above formulas to fit experimentally measurable transient-absorption spectra. In particular, one can fit probe-pump spectra to quantify $U_{\mathrm{pu},11}\,U_{\mathrm{pu},22}^*$, $U_{\mathrm{pu},11}\,U_{\mathrm{pu},33}^*$, $U_{\mathrm{pu},11}\,U_{\mathrm{pu},32}^*$ and $U_{\mathrm{pu},11}\,U_{\mathrm{pu},23}^*$, and the common multiplication factor $K$. Once $K$ is known, it can be employed to fit pump-probe spectra, and thereby extract $|U_{\mathrm{pu},11}|^2$, $|U_{\mathrm{pu},21}|^2$, $|U_{\mathrm{pu},31}|^2$, and $(U_{\mathrm{pu,}31} \,U^*_{\mathrm{pu,}21})$. 

All elements of the interaction operator $\hat{U}_{\mathrm{pu}}$ which can be retrieved from probe-pump spectra are inferred from product terms $U_{\mathrm{pu},11} U_{\mathrm{pu}, ij}^*$, with $|U_{\mathrm{pu},11}|$ coming from pump-probe spectra. This has two consequences. Firstly, the phase $\beta$ of $U_{\mathrm{pu},11} = |U_{\mathrm{pu},11}|\eu^{\uimm\beta}$ cannot be accessed, resulting in SFI matrix elements 
\begin{equation}
U_{\mathrm{pu}, ij} = \frac{(U_{\mathrm{pu},11} U_{\mathrm{pu}, ij}^*)^*}{|U_{\mathrm{pu},11}|}\eu^{\uimm\beta},
\label{eq:ratio}
\end{equation}
known up to this common phase $\beta$. However, this is not a limitation, since it only implies that the final state can be measured and controlled up to a nonrelevant phase term, with access to the information about relevant relative phases. Secondly, when $|U_{\mathrm{pu}, 11}|$ is very close to $0$, small uncertainties in its reconstructed value are amplified when the division in Eq.~(\ref{eq:ratio}) is performed to retrieve the remaining matrix elements. For a certain range of pump-pulse intensities, we verified that this is indeed the main source of uncertainty in the extraction of $\hat{U}_{\mathrm{pu}}$, yet always below the relative level of $8\%$.

In a noncollinear geometry, the remaining pump-probe fitting parameter $(U_{\mathrm{pu,}31} \,U^*_{\mathrm{pu,}21})$ can be used to quantify the relative phase $\arg{(U_{\mathrm{pu,}31})} - \arg{(U_{\mathrm{pu,}21})}$, but not the absolute phases of $U_{\mathrm{pu,}21}$ and $U_{\mathrm{pu,}31}$---at least up to the same common phase $\beta$ we introduced before. However, this can be easily circumvented by directly observing the absorption spectrum of a single intense pump pulse. The associated spectral lines are given by
\begin{equation}
\mathcal{S}_{\mathrm{fit}}(\omega) 
\propto\,-\omega\Imm\biggl\{\mathcal{M} + \sum_{k=2}^3 D_{1k}^* \int_{0}^{\infty}\rho_{\mathrm{fit},1k}(t)\,\eu^{-\uimm\omega t}\,\diff t  \biggr\}\,,
\label{eq:Sexpnopr}
\end{equation}
where $\mathcal{M}$ is a fitting parameter modeling the broadband (and hence constant for small frequency intervals) Fourier transform of $\rho^{(3)}_{1k}(t)$ in the interval $[-T/2,\,T/2]$, i.e., in the presence of the pump pulse. The remaining integral 
\begin{equation}
\begin{split}
&\,\int_{0}^{\infty}\rho_{\mathrm{fit},1k}(t)\,\eu^{-\uimm\omega t }\,\diff t \\
= &\,U_{\mathrm{pu},11}\,U_{\mathrm{pu},k1}^*\int_{0}^{\infty} \,\eu^{-\uimm(\omega - \omega_{k1})t} \,\eu^{-\frac{\gamma_k}{2}t}\,\diff t \\
= &\,U_{\mathrm{pu},11}\,U_{\mathrm{pu},k1}^*\,\frac{1}{\uimm(\omega - \omega_{k1}) + \frac{\gamma_k}{2}}
\end{split}
\label{eq:integral1pulse}
\end{equation}
exploits the effective evolution of the system, given by
\begin{equation}
|\psi_{\mathrm{fit}}(t)\rangle = \left\{
\begin{aligned}
&|\psi_0\rangle, &\text{if $t<0$},\\
&\hat{V}(t)\hat{U}_{\mathrm{pu}}(I)|\psi_0\rangle, &\text{if $t>0$}.
\end{aligned}
\right.
\label{eq:evolutionnoprobe}
\end{equation}
Up to the same common phase $\beta$ already mentioned above, this allows one to retrieve the phases of $U_{\mathrm{pu},21}$ and $U_{\mathrm{pu},31}$, complementing the amplitude information accessible by fitting pump-probe spectra.

In order to independently extract $U_{\mathrm{pu},12}$ and $U_{\mathrm{pu},13}$ in a noncollinear geometry, we show that, when the envelope function $f(t) = f(-t)$ is a symmetric function of time, then the associated interaction operator $\hat{U} = \hat{V}(-T/2)\,\hat{\mathcal{U}}_0(T/2, -T/2)\,\hat{V}^{-1}(T/2)$ is a symmetric matrix. We recall that the time evolution of $\hat{\mathcal{U}}_0(t,t_0)$ is calculated for a pulse with $\phi = 0$ and $t_{\mathrm{c}} = 0$, such that
\begin{equation}
\begin{split}
\frac{\diff \hat{\mathcal{U}}_0(t,t_0)}{\diff t} &= [\hat{\varLambda}^{(3)}(t)]\daga\,\hat{M}^{(3)}(t)\,\hat{\varLambda}^{(3)}(t)\,\hat{\mathcal{U}}(t,t_0),\\
\hat{\mathcal{U}}(t_0,t_0) &= \hat{I},
\end{split}
\label{eq:timeevolution3U}
\end{equation}
compare this with Eq.~(\ref{eq:timeevolution5}). 
The assumption of a symmetric envelope, $f(t) = f(-t)$, implies that also $ \hat{M}^{(3)}(t)$ is symmetric in time. 

We thus introduce the operator $\hat{\mathcal{Z}}(t,t_0) = \hat{\mathcal{U}}_0(-t, -t_0)$, which is solution of the differential equation
\begin{equation}
\begin{split}
\frac{\diff \hat{\mathcal{Z}}(t,t_0)}{\diff t} &= -\hat{\varLambda}^{(3)}(t)\,\hat{M}^{(3)}(t)\,[\hat{\varLambda}^{(3)}(t)]^{-1}\,\hat{\mathcal{Z}}(t,t_0),\\
\hat{\mathcal{Z}}(t_0,t_0) &= \hat{I},
\end{split}
\label{eq:evolutionZ}
\end{equation}
where $\hat{\varLambda}^{(3)}(-t) = [\hat{\varLambda}^{(3)}(t)]^{-1}$ and $\hat{M}^{(3)}(-t) = \hat{M}^{(3)}(t)$. Both $\hat{\varLambda}^{(3)}(t) = [\hat{\varLambda}^{(3)}(t)]^{\mathrm{T}}$ and $\hat{M}^{(3)}(t) = [\hat{M}^{(3)}(t)]^{\mathrm{T}}$ are symmetric matrices. As a result,
\begin{equation}
\begin{split}
\frac{\diff \hat{\mathcal{Z}}^{-1}(t,t_0)}{\diff t} 
&= \hat{\mathcal{Z}}^{-1}(t,t_0)\hat{\varLambda}^{(3)}(t)\,\hat{M}^{(3)}(t)\,[\hat{\varLambda}^{(3)}(t)]^{-1},\\
\hat{\mathcal{Z}}^{-1}(t_0,t_0) & = \hat{I},
\end{split}
\label{eq:evolutionZ-1}
\end{equation}
and 
\begin{equation}
\begin{split}
\frac{\diff (\hat{\mathcal{Z}}^{-1})^{\mathrm{T}}(t,t_0)}{\diff t} 
&\,=[\hat{\varLambda}^{(3)}(t)]^{-1}\,\hat{M}^{(3)}(t)\,\hat{\varLambda}^{(3)}(t)\,(\hat{\mathcal{Z}}^{-1})^{\mathrm{T}}(t,t_0),\\
(\hat{\mathcal{Z}}^{-1})^{\mathrm{T}}(t_0,t_0) &\,= \hat{I}.
\end{split}
\label{eq:evolutionZ-1T}
\end{equation}
Since this equation corresponds to Eq.~(\ref{eq:timeevolution3U}), the evolution operators $(\hat{\mathcal{Z}}^{-1})^{\mathrm{T}}(t,t_0)$ and $\hat{\mathcal{U}}_0(t, t_0)$ are solution of the same differential equations and are therefore identical, and hence $\hat{\mathcal{U}}_0(T/2,0) = \hat{\mathcal{U}}_0^{-1}(0, T/2) = \hat{\mathcal{Z}}^{-1}(0,-T/2) = \hat{\mathcal{U}}_0^{\mathrm{T}}(0,-T/2)$.
As a result,
\begin{equation}
\begin{split}
\hat{\mathcal{U}}_0(T/2,-T/2) &=  \hat{\mathcal{U}}_0(T/2,0)\,\hat{\mathcal{U}}_0(0,-T/2)\\
&=\hat{\mathcal{U}}_0^{\mathrm{T}}(0,-T/2) \,\hat{\mathcal{U}}_0(0,-T/2),
\end{split}
\end{equation}
and
\begin{equation}
\begin{split}
\hat{\mathcal{U}}_0^{\mathrm{T}}(T/2,-T/2) &= \hat{\mathcal{U}}_0^{\mathrm{T}}(0,-T/2) \,\hat{\mathcal{U}}_0(0,-T/2) \\
&= \hat{\mathcal{U}}_0(T/2,-T/2),
\end{split}
\end{equation}
i.e., $\hat{\mathcal{U}}_0(T/2,-T/2) $ is a symmetric matrix. Since $\hat{V}(t)$ is a diagonal (and hence symmetric) matrix, it follows that $\hat{U} = \hat{V}(-T/2)\,\hat{\mathcal{U}}_0(T/2, -T/2)\,\hat{V}^{-1}(T/2)$ is symmetric. 

We hence use the above result for the operator $\hat{U}_{\mathrm{pu}}$ reconstructed from strong-field TAS, and fix $U_{\mathrm{pu},1k} = U_{\mathrm{pu},k1}$. We stress that this is not a disadvantage of the operator-reconstruction scheme proposed here. As shown by Eq.~(\ref{eq:spprpu2ndcomponent}), an independent extraction of $U_{\mathrm{pu},12}$ and $U_{\mathrm{pu},13}$, without additional assumptions on the envelope $f(t)$, is possible in a collinear geometry.

{
\section{Density-matrix formulation}
\label{Density-matrix formulation}

The equations of motion have been investigated thus far in the framework of the Schr\"odinger equation, assuming a decay-rate model with a complex-valued atomic structure Hamiltonian. For our implementation with Rb atoms, this was a proper treatment, in the light of the very different time scales in which the strong-field dynamics and the relaxation of the system take place. 
For more general cases, however, a density-matrix formulation of our operator-reconstruction and quantum-control scheme is possible, as we show in the following. In this section, we will focus on the case of a three-level system, although the equations can be easily generalized to higher numbers of levels if necessary.

We describe the system in terms of a $3\times 3$ density matrix $\hat{\rho}(t)$, with a diagonal element $\rho_{ii}(t)$ associated with the population of the $i$th level, and off-diagonal matrix elements for the atomic coherences. A density-matrix formulation may be necessary, e.g., to explicitly account for the spontaneous decay of the two excited levels $2$ and $3$ into the ground state $1$, at rates respectively given by $\gamma_{22}$ and $\gamma_{33}$. Similarly, it is necessary to account for additional decoherence processes affecting the atomic coherences $\rho_{12}$, $\rho_{13}$, and $\rho_{23}$. In this Section, this is modeled by the decoherence rates $\gamma_{21}$, $\gamma_{31}$, and $\gamma_{23}$, respectively. The evolution of the density matrix $\hat{\rho}(t)$ is then given by the master equation 
\begin{equation}
\frac{\diff \hat{\rho}(t)}{\diff t} = \uimm [\hat{\rho}(t),\,\hat{H}_0 + \hat{H}_{\mathrm{int}}] + \mathcal{L}[\hat{\rho}(t)],
\end{equation}
with the atomic structure Hamiltonian now given by $\hat{H}_ 0  = \sum \omega_i |i\rangle\langle i|$ and the superoperator $\mathcal{L}[\hat{\rho}(t)]$ describing dissipation. In the absence of external fields, the equations of motion satisfied by the elements of the density matrix can hence be written as:
\begin{equation}
\begin{split}
\frac{\diff \rho_{11}}{\diff t} &= \gamma_{22}\rho_{22} + \gamma_{33}\rho_{33},\\
\frac{\diff \rho_{kk}}{\diff t} &= -\gamma_{kk}\rho_{kk},\\
\frac{\diff \rho_{1k}}{\diff t} &=  - \gamma_{k1}\rho_{1k} + \uimm \omega_{k1} \rho_{1k},\\
\frac{\diff \rho_{23}}{\diff t} &=  - \gamma_{23}\rho_{23} + \uimm \omega_{32} \rho_{23},
\end{split}
\end{equation}
with $k\in\{2,3\}$ and $\rho_{ij} = \rho_{ji}^*$. The explicit solution reads
\begin{equation}
\begin{split}
\rho_{11}(t) &= \rho_{11}(0) + \rho_{22}(0)\left(1 - \eu^{-\gamma_{22}t} \right) \\
&\ \ \ + \rho_{33}(0)\left(1 - \eu^{-\gamma_{33}t} \right),\\
 \rho_{kk}(t) &= \rho_{kk}(0) \eu^{-\gamma_{kk}t},\\
 \rho_{1k}(t) &=  \rho_{1k}(0) \eu^{ - \gamma_{k1}t}  \eu^{\uimm \omega_{k1} t},\\
 \rho_{23}(t) &=  \rho_{23}(0) \eu^{ - \gamma_{23}t}  \eu^{\uimm \omega_{32} t},
\end{split}
\end{equation}
summarized by the elements of the operator $\boldsymbol{X}(t)$ via
\begin{equation}
\rho_{ij}(t) = \sum_{i'j'}\boldsymbol{X}_{ij,i'j'}(t)\rho_{i'j'}(0).
\end{equation}

In analogy to Eq.~(1) in the article, we can introduce an operator $\boldsymbol{Y}$ describing the effect of ultrashort pump and probe pulses on the density matrix of the system as an instantaneous interaction, i.e.,
\begin{equation}
\rho^+ = \boldsymbol{Y}(I)\rho^-.
\end{equation}
For ultrashort pulses, acting on a time scale much shorter than the time scale of the decay processes, we can assume that $\boldsymbol{Y}(I) = \hat{U}(I)\otimes \hat{U}^*(I)$, i.e., $\boldsymbol{Y}_{ij,i'j'} = U_{ii'}\,U_{jj'}^*$. In contrast to the previous Sections, however, here we fully account for the relaxation of the system during its free evolution in terms of the operator $\boldsymbol{X}(t)$. A similar approach can be used to model quantum systems also in the presence of a more structured reservoir, provided that the free evolution of the system is known and can be used to define the operator $\boldsymbol{X}(t)$.

For a TAS experiment in a probe-pump setup, we can describe the effective evolution of the time-delay-dependent density matrix $\hat{\rho}_{\mathrm{fit}}(t,\tau)$ from the effective initial state $\rho_0 = |1\rangle\langle 1|$ as
\begin{equation}
\hat{\rho}_{\mathrm{fit}}(t,\tau) = \left\{
\begin{aligned}
&\hat{\rho}_0, &\text{if $t<\tau$},\\
&\boldsymbol{X}(t-\tau)\boldsymbol{Y}_{\mathrm{pr}}\hat{\rho}_0, &\text{if $\tau<t<0$},\\
&\boldsymbol{X}(t)\boldsymbol{Y}_{\mathrm{pu}}(I)\,\boldsymbol{X}(-\tau)\,\boldsymbol{Y}_{\mathrm{pr}}\hat{\rho}_0, &\text{if $t>0$}.
\end{aligned}
\right.
\end{equation}
By inserting $\rho_{\mathrm{fit},ij}(t,\tau)$ into Eq.~(\ref{eq:spectrum3}), an analytical interpretation model for the probe-pump spectrum is obtained consisting also in this case of two integrals. The first integral reads
\begin{equation}
\begin{split}
&\,\int_{\tau}^{0}\rho_{\mathrm{fit},1k}(t,\tau)\,\eu^{-\uimm\omega (t-\tau)}\,\diff t \\
=&\int_{\tau}^{0} \sum_{i,j=1}^3\boldsymbol{X}_{1k,ij}(t-\tau)\,\boldsymbol{Y}_{\mathrm{pr};ij,11} \,\eu^{-\uimm\omega (t-\tau)}\,\diff t \\
= &\, -\uimm\frac{\vartheta_k}{2}\,\frac{1-\eu^{\uimm(\omega - \omega_{k1}) \tau}\,\eu^{\gamma_{k1}\tau}}{\uimm(\omega - \omega_{k1}) + \gamma_{k1}},
\end{split}
\end{equation}
where we have used the fact that $\boldsymbol{X}_{1k,ij}(t) = \eu^{-\gamma_{k1}t}\,\eu^{\uimm\omega_{k1}t}\,\delta_{1i}\,\delta_{kj}$. The second integral is similarly given by
\begin{equation}
\begin{split}
&\,\int_{0}^{\infty}\rho_{\mathrm{fit},1k}(t,\tau)\,\eu^{-\uimm\omega (t-\tau)}\,\diff t \\
=&\,\sum_{i,j =1}^3\sum_{i',j' =1}^3\sum_{i'',j'' =1}^3 \boldsymbol{Y}_{\mathrm{pu};ij,i'j'}\,\boldsymbol{X}_{i'j',i''j''}(-\tau)\,\boldsymbol{Y}_{\mathrm{pr},i''j'',11}\\
&\times\,\int_{0}^{\infty} \boldsymbol{X}_{1k,ij}(t)\,\eu^{-\uimm\omega (t-\tau)} \,\diff t \\
= &\,\frac{1}{\,\uimm(\omega - \omega_{k1}) + \gamma_{k1}}\sum_{i',j' =1}^3\sum_{i'',j'' =1}^3 U_{\mathrm{pu},1i'}\,U_{\mathrm{pu},kj'}^*\,\\
&\ \ \times\,\left[\boldsymbol{X}_{i'j',i''j''}(-\tau)\,\eu^{\uimm\omega\tau}\right]\, U_{\mathrm{pr},i''1}\,U_{\mathrm{pr},j''1}^*. 
\end{split}
\end{equation}
By including explicitly the value of $\boldsymbol{X}_{i'j',i''j''}(-\tau)$, one obtains an equation in which, in contrast to Eq.~(\ref{eq:spprpu2ndcomponent}), different decay rates for coherences and populations are present and additional terms appear due to the spontaneous decay from the excited states to the ground state. 
In order to model spectra from a noncollinear geometry, fast oscillating terms due to $[\boldsymbol{X}_{i'j',i''j''}(-\tau)\,\eu^{\uimm\omega\tau}]$ can be eliminated, resulting in the average spectrum 
\begin{equation}
\begin{split}
&\langle \mathcal{S}_{\mathrm{fit}}(\omega,\tau,U_{\mathrm{pu},ij})\rangle_{\tau} \\
\propto &-\omega\Imm\biggl\{\sum_{k=2}^3 D_{1k}^* \,\frac{1}{\uimm(\omega - \omega_{k1}) + \gamma_{k1}}\\
&\ \ \ \ \ \ \ \biggl[(-\uimm{\vartheta_k}/{2})\,\bigl(1-\eu^{\uimm(\omega - \omega_{k1}) \tau}\,\eu^{\gamma_{k1}\tau}\bigr)\\
&\ \ \ \ \ \ \ \ + U_{\mathrm{pu},11}\,U_{\mathrm{pu},k2}^*\, (-\uimm{\vartheta_2}/{2})\,\eu^{\uimm(\omega-\omega_{21})\tau}\,\eu^{\gamma_{21}\tau}\\
&\ \ \ \ \ \ \ \ \ +U_{\mathrm{pu},11}\,U_{\mathrm{pu},k3}^*\, (-\uimm{\vartheta_3}/{2})\, \eu^{\uimm(\omega-\omega_{31})\tau}\,\eu^{\gamma_{31}\tau}  \biggr]\biggr\}.
\end{split}
\end{equation}

In a pump-probe setup, the same formalism can be used, resulting in an effective evolution given by
\begin{equation}
\hat{\rho}_{\mathrm{fit}}(t,\tau) = \left\{
\begin{aligned}
&\hat{\rho}_0, &\text{if $t<0$},\\
&\boldsymbol{X}(t)\,\boldsymbol{Y}_{\mathrm{pu}}(I)\,\hat{\rho}_0, &\text{if $0<t<\tau$},\\
&\boldsymbol{X}(t-\tau)\,\boldsymbol{Y}_{\mathrm{pr}}\,\boldsymbol{X}(\tau)\,\boldsymbol{Y}_{\mathrm{pu}}(I)\,\hat{\rho}_0, &\text{if $t>\tau$}.
\end{aligned}
\right.
\end{equation}
By following the same procedure outlined for the probe-pump case, and after removing the fast oscillating terms, one obtains the average spectrum given by
\begin{equation}
\begin{split}
&\langle \mathcal{S}_{\mathrm{fit}}(\omega,\tau, U_{\mathrm{pu},ij})\rangle_{\tau} \\
\propto &-\omega\Imm\biggl\{\frac{D_{12}^*}{\uimm(\omega - \omega_{21}) + \gamma_{21}}\biggl[ \Bigl( -\uimm\frac{\vartheta_{2}}{2}\, |U_{\mathrm{pu,}11}|^2\\
&\ \ \ \ \ \ \ \ \  +\uimm\frac{\vartheta_{2}}{2}\, |U_{\mathrm{pu,}21}|^2 \,(2\eu^{-\gamma_{22}\tau}-1) \Bigr)\\
&\ \ \ \ \ \ \ \ \  +\uimm\frac{\vartheta_{3}}{2}\,U_{\mathrm{pu,}31} \,U^*_{\mathrm{pu,}21} \,\eu^{-\uimm\omega_{32}\tau}\,\eu^{-\gamma_{23}\tau} \biggr]\\
&\ \ \ \ \  + \frac{D_{13}^*}{\uimm(\omega - \omega_{31}) + \gamma_{31}}\biggl[ \Bigl( -\uimm\frac{\vartheta_{3}}{2}\, |U_{\mathrm{pu,}11}|^2\\
&\ \ \ \ \ \ \ \ \  +\uimm\frac{\vartheta_{3}}{2}\, |U_{\mathrm{pu,}31}|^2 \,(2\eu^{-\gamma_{33}\tau}-1)\Bigr)\\
&\ \ \ \ \ \ \ \ \  +\uimm\frac{\vartheta_{2}}{2}\, U_{\mathrm{pu,}21}\,U^*_{\mathrm{pu,}31}\,\eu^{\uimm\omega_{32}\tau}\,\eu^{-\gamma_{23}\tau}\biggr]\biggr\}.
\end{split}
\end{equation}
Also in this case, in contrast to Eq.~(\ref{eq:Sexppupr}), we notice different decay rates and the presence of additional terms to account for the decay into the ground state 1.

This model can be employed to fit experimental results and extract interaction operators $\boldsymbol{Y}$ in those cases in which a density-matrix description of the free evolution of the system is required. Once SFI operators are reconstructed, then the sequence of pulses constituting our control method can be modeled as
\begin{equation}
\hat{\rho}_{N_{\mathrm{p}}} = \tilde{\boldsymbol{Y}}_{N_{\mathrm{p}}} \ldots \boldsymbol{X}(\bar{\tau}_m)\tilde{\boldsymbol{Y}}_{m}\ldots \boldsymbol{X}(\bar{\tau}_1)\tilde{\boldsymbol{Y}}_1\hat{\rho}_0,
\end{equation}
generalizing Eq.~(2) in the article to the case when the system's dynamics have to be described by a density matrix.

}